\documentclass[%
 reprint,
 superscriptaddress,
nofootinbib,
 amsmath,amssymb,
 aps,
 pra,
]{revtex4-2}
\usepackage{amsmath,amsfonts}
\usepackage{algorithmic}
\usepackage{array}
\usepackage[caption=false,font=normalsize,labelfont=sf,textfont=sf]{subfig}
\usepackage{textcomp}
\usepackage{url}
\usepackage{hyperref}
\usepackage{verbatim}
\usepackage{graphicx}
\graphicspath{{./Figures/}}
\def\BibTeX{{\rm B\kern-.05em{\sc i\kern-.025em b}\kern-.08em
    T\kern-.1667em\lower.7ex\hbox{E}\kern-.125emX}}
\usepackage{balance}
\usepackage{dcolumn}
\usepackage{siunitx}
\usepackage{bm}
\usepackage{bbold}
\usepackage{braket}
\usepackage{esdiff}
\usepackage[skins,theorems]{tcolorbox}
\usepackage{color}
\usepackage{blindtext}
\usepackage{subfiles} 

\begin{document}


\preprint{APS/123-QED}

\title{Pulse shape optimization against Doppler shifts and delays in optical quantum communication
}

\author{Emanuel Schlake}
\email{emanuel.schlake@zarm.uni-bremen.de}
\affiliation{ZARM, University of Bremen, 28359 Bremen, Germany}%
\affiliation{Gauss-Olbers
Space Technology
Transfer Center, University of Bremen, 28359 Bremen, Germany}
\affiliation{Department of Communications Engineering, University of Bremen, 28359 Bremen, Germany}
\author{Roy Barzel}
\affiliation{ZARM, University of Bremen, 28359 Bremen, Germany}%
\affiliation{Gauss-Olbers
Space Technology
Transfer Center, University of Bremen, 28359 Bremen, Germany}
\author{Dennis Rätzel}
\email{dennis.raetzel@zarm.uni-bremen.de}
\affiliation{ZARM, University of Bremen, 28359 Bremen, Germany}%
\affiliation{Gauss-Olbers
Space Technology
Transfer Center, University of Bremen, 28359 Bremen, Germany}
\author{Claus Lämmerzahl} 
\affiliation{ZARM, University of Bremen, 28359 Bremen, Germany}%
\affiliation{Gauss-Olbers
Space Technology
Transfer Center, University of Bremen, 28359 Bremen, Germany}

\date{\today}


\clearpage
\newpage

\begin{abstract}
    High relative velocities and large distances in space-based quantum communication with satellites in lower earth orbits can lead to significant Doppler shifts and delays of the signal impairing the achievable performance if uncorrected.
    We analyze the influence of systematic and stochastic Doppler shift and delay in the specific case of a continuous variable quantum key distribution (CV-QKD) protocol and identify the generalized correlation function, the ambiguity function,
    as a decisive measure of performance loss. Investigating the generalized correlations as well as private capacity bounds for specific choices of spectral amplitude shape (Gaussian, single- and double-sided Lorentzian), we find that this choice has a significant impact on the robustness of the quantum communication protocol to spectral and temporal synchronization errors. 
    We conclude that optimizing the pulse shape can be a building block in the resilient design of quantum network infrastructure.
\end{abstract}

\maketitle

\section{Introduction \label{sec:introduction}}

The second quantum revolution \cite{dowlingQuantumTechnologySecond2003, bellSpeakableUnspeakableQuantum2004} is bringing physical concepts into technological realization. Such are quantum networks \cite{bassoliQuantumCommunicationNetworks2021}, made up of optical quantum channels and nodes, eventually culminating in the quantum internet \cite{kimbleQuantumInternet2008, azumaQuantumRepeatersQuantumInternet2023, Rohde_Internet_2021}.
Such a network would interconnect various domains of quantum technology, enabling the distribution of quantum resources such as entanglement, quantum computing, sensor networks, and large-scale secure communication channels \cite{awschalom_internet}. Because quantum systems are inherently prone to decoherence their channels must be protected from the environment.

A fiber-based approach to such networks is severely limited in distance as the transmittance of optical fibers decreases exponentially with its length. Together with the fundamental rate-loss scaling of channel capacities \cite{pirandola2017fundamental}, this makes optical fibers alone incompatible with long-distance, point-to-point transmissions. Integration of quantum repeaters \cite{briegel1998quantum,azumaQuantumRepeatersQuantumInternet2023} into the network would alleviate this issue, however, for a large-scale network, a large number of repeaters would be required \cite{sidhu2021advances}, making a pure fiber-based realization unfeasible in the foreseeable future.
Therefore, the local fiber-based networks will likely be supplemented by long-distance space-based networks, opening the realm of satellite-based quantum communication \cite{sidhu2021advances, deforgesdeparnySatellitebasedQuantumInformation2023, khatriSpookyActionGlobal2021, BELENCHIA20221_Quantum_Physics_in_Space}.
Its feasibility has already been demonstrated with the Micius satellite \cite{ren2017ground, yin2017satellite, yin2017satellite2, chenIntegratedSpacetogroundQuantum2021} and further ambitious plans such as the QEYSSat \cite{QEYSSAT,jenneweinQuantumCommunicationsSatellites2018, jenneweinQEYSSatWhitePaper2024} and QUBE \cite{QUBE} are developing.

A major difficulty in satellite-based communication, especially with lower-earth orbit (LEO) satellites, is the large relative velocities and distances between satellites and ground stations. The 
relative velocities translate to a Doppler shift of the transmitted optical signals, deforming the signal spectrum
as well as deteriorating the bit rates \cite{zhaoEffectDopplerShift2010, wangFeasibilitySpacebasedMeasurementdeviceindependent2021, gaoPerformanceAnalysisDownlink2020}.
The large distances complicate the precise synchronization between communicating parties needed to coordinate their communication effort.
The optical domain of quantum communication amplifies these difficulties twofold; first, the Doppler effect is proportional to the signal frequency and therefore especially significant in the optical (THz) regime, second quantum optical communication requires coherent detection techniques which are inherently more sensitive to synchronization errors.
Furthermore, any practical quantum network will most likely contain quantum memories as vital elements as they enable the storage of quantum information and can help increase communication rates.
In space-based applications, the memories are typically required to be long-lived \cite{gundoganProposalSpaceborneQuantum2021, gundoganTimedelayedSingleSatellite2024} and therefore operate with spectrally narrow signals making them vulnerable to Doppler shifts.
There have been efforts to characterize the Doppler shift in LEO constellations \cite{ali1998doppler, yangDopplerCharacterizationLaser2009a}, to estimate their effect \cite{vilarAnalysisCorrectionTechniques1991, amiriAccurateDopplerFrequency2007, zhaoEffectDopplerShift2010} and to compensate it \cite{youAdaptiveCompensationMethod2000, hongDopplerAnalysisCompensation2022a, andoCoherentHomodyneReceiver2011, andoHomodyneBPSKReceiver2011}.
While certainly fruitful, we propose integrating inherently resilient designs into the communication systems alongside these compensation efforts.
In this article, we investigate the effect of the choice of spectral amplitude waveform, or equivalently, the pulse shape of the signal carrier, on channel capacities as performance measures of optical quantum communication with a focus on continuous-variable quantum key distribution (CV-QKD).
Although satellite-based CV-QKD protocols have been proposed and their feasibility investigated \cite{dequal2021feasibility}, the effect of Doppler shift in these scenarios has not been analyzed quantitatively.
While a broader signal bandwidth naturally mitigates the effects of Doppler shifts, it shortens the duration of the pulse making it vulnerable to delays. Mathematically this follows from the scaling property of the Fourier transformation. However, even at equal bandwidth, the exact pulse shape (or spectral amplitude) changes the effects of Doppler shifts and delays.
Depending on how well either Doppler shift or synchronization can be controlled, the choice of pulse shape can help to mitigate these.

The work is organized as follows: In Section \ref{sec:photonic_wavepackets}, we introduce the formalism of continuum quantum states of light and their spectral amplitude shapes. We then determine the exact deformation under Doppler shift and delay in Section \ref{sec:deformation}.
In Section \ref{sec:overlap}, we introduce the overlap integral and from this derive the ambiguity function as a general measure for correlation and mode match in the presence of delay and Doppler shift. In Section \ref{sec:cvqkd}, we derive in detail how mode mismatch enters into an exemplary CV-QKD protocol with homodyne detection.

We show that the quantum channel due to delay and Doppler shift becomes a lossy dephasing channel and discuss channel capacities.
In Section \ref{sec:pulse_shape_optimization}, we consider explicitly the Gaussian and single- and double-sided Lorentzian spectral amplitude functions and discuss their robustness under delay and Doppler shift by investigating their ambiguity function and the private capacity bound for the lossy dephasing channel. We conclude in Section \ref{sec:conclusion}.


\section{Quantum states of light and deformation of spectral profiles}\label{sec:statesoflight}
To investigate the spectral properties of quantum states of light we assume a quantized electric field. Following \cite{blowContinuumFieldsQuantum1990, shapiroQuantumTheoryOptical2009}, we can simplify the analysis to essentially one dimension since we are focusing on free-space propagation.
Fixing the propagation direction to the $z$-axis and fixing the polarization to the $x$-axis, we can write the electric field operator for continuum fields as a decomposition in continuous frequency modes $\omega \in \mathbb{R}_+$ given by the following scalar function 
\begin{equation}
    E^{+}(z,t) = i \int_0^\infty d\omega \kappa_\omega a_\omega e^{-i\omega (t-\frac{z}{c})},
\end{equation}
where $\kappa_\omega = \sqrt{\frac{\hbar \omega}{4\pi \epsilon_0 c A}}$ is the electrical field strength per photon in the mode of angular frequency $\omega$,
$A$ is the transversal beam width, $c$ is the speed of light and $\epsilon_0$ is the vacuum permittivity. The annihilation and creation operators fulfill the usual commutation relations
\begin{equation}
    [a_\omega, a_{\omega'}^\dagger] = \delta(\omega-\omega').
\end{equation}

\subsection{Photonic wavepackets}\label{sec:photonic_wavepackets}

To describe spectra instead of single frequencies, we now define the creation operator for some arbitrary photon wavepacket shape which is given by some normalized spectral distribution $F(\omega_A)$
\begin{equation}
    a^\dagger_{F} = \int_0^\infty d\omega F(\omega)a^\dagger(\omega)
\end{equation}
The creation and annihilation operators
fulfill the usual commutator relation
\begin{equation}
    [a_{F}, a^\dag_{F}] = 1.
\end{equation}

The commutator of different wavepacket creation and annihilation operators is given by the overlap of their spectral distributions
\begin{equation}
    [a_{F}, a^\dag_{F'}] = \int d\omega F^*(\omega) {F'(\omega)}.
\end{equation}
One can then further introduce a non-continuous set of orthonormal functions $\{\xi_j\}$ such that the corresponding creation operators $a_{\xi_j}$ commute as $[a_{\xi_i}, a^\dag_{\xi_j}] = \delta_{ij}$. This enables a convenient decomposition into orthogonal modes.
Such a set of orthonormal functions can be constructed from any given spectrum
$F = \xi_0$ via Gram-Schmidt orthogonalization.
For example, if $F$ was a Gaussian, the orthonormal functions would be the Hermite-Gaussian functions \cite{blowContinuumFieldsQuantum1990}.

From these photon wavepacket creation operators, we can construct more complex field states \cite{blowContinuumFieldsQuantum1990}.
Multimode Fock states are constructed as usual
\begin{equation}
    \label{eq:Fock}
    \Ket{\{n_i\}} = \prod_j \frac{(a^\dagger_{\xi_j})^{n_j}}{\sqrt{n_j!}} \Ket{0}.
\end{equation}
Continuum mode coherent states are constructed from the generalized displacement operator for some spectral amplitude $F(\omega)$
\begin{equation}
    \label{eq:cont_coherent}
\begin{aligned}
    \Ket{\alpha}_F &= \bigotimes_\omega \Ket{\alpha F(\omega)} =
    e^{\int d\omega \left(
    \alpha F(\omega) a^\dagger(\omega) - \alpha^*F^*(\omega) a_\omega \right)} \Ket{0},
\end{aligned}
\end{equation}
where $\alpha = |\alpha|e^{i\theta}$ is the coherent state's complex amplitude such that $|\alpha|^2 = \braket{n}$ is the mean total photon number.
Introducing a complete set of orthonormal functions $\{\xi_i\}$,  we can again perform the following mode decomposition
\begin{equation}
     \Ket{\alpha}_F =
    e^{\sum_i \left(c_i a^\dagger_{\xi_i} - c^*_i a_{\xi_i} \right) } \Ket{0}
    = \prod_i \Ket{c_i \xi_i},
\end{equation}
where $c_i = \alpha \int d\omega  F(\omega) \xi^*_i(\omega)$ are the Gram-Schmidt coefficients.
Squeezed states are analogously constructed from a generalized squeezing operator, as done in e.g. \cite{blowContinuumFieldsQuantum1990}.

\subsection{Signal deformation}\label{sec:deformation}
As the signal propagates between two parties in relative motion (e.g. satellites), at reception it will be distorted with respect to the expected signal.
Possible reasons for this are noise sources at the emitter's and receiver's end as well as the propagation through a possibly inhomogeneous medium.
The relative motion of the parties distorts the signal spectrally and temporally via the Doppler shift.
For free space propagation, the relativistic effects of gravitational redshift and corrections to the Doppler shift are noteworthy.
\footnote{The full treatment would of course require solving the actual wave equation.}

In the following, we focus on free space propagation taking into account only Doppler shifts and delays, applying our previously introduced one-dimensional formalism for spectral amplitudes describing photon wavepackets.
An application where this is highly relevant is satellite communication with lower earth orbits where the high relative velocities cause relative shifts of the order of $10^{-5} = 10 \mathrm{ppm}$ \cite{ali1998doppler}.
Depending on the constellation, the rate of change of the Doppler shift (Doppler rate) might also vary quickly such that a residual shift will remain (e.g. \cite{vilarAnalysisCorrectionTechniques1991}) even when compensation is applied. In Section \ref{sec:quantumchannelcapacity}, we will discuss residual errors.

The general relative shift (or rather stretch) in frequency $z$ between an emitter and receiver is 
\begin{equation}\label{eq:freq_shift}
    z = \frac{\omega-\omega_\mathrm{rec}}{\omega_\mathrm{rec}} = \frac{\Delta \omega}{\omega_\mathrm{rec}}
    \quad \Leftrightarrow \quad
    \omega_\mathrm{rec} = \frac{\omega}{1+z} ,
\end{equation}
here $\omega$ and $\omega_\mathrm{rec}$ are the emitted and received frequency, respectively. This is a general expression accounting for various sources of frequency shift such as the classical and relativistic Doppler shift and gravitational redshift. 

For monochromatic narrowband 
signals, the Doppler effect results in a simple frequency shift. In spectral distributions, each frequency experiences a different shift, causing a stretch or compression of the spectrum.
The argument of the spectral amplitude changes according to \eqref{eq:freq_shift}, and the received (normalized) amplitude $F_\mathrm{rec}(\omega)$ can then be expressed in terms of the emitted amplitude 
$F(\omega)$ as
\begin{equation}
    F_\mathrm{rec}(\omega) 
     = \sqrt{1+z}F\left((1+z)\omega\right) 
    \label{eq:dist_shift}
\end{equation}
where the factor $\sqrt{1+z}$ preserves the normalization.

The normalized signal shape in the temporal domain $A(t)$ is related to the spectral amplitude by a Fourier transformation,
\begin{equation}
A(t) = \frac{1}{\sqrt{2\pi}}\int d\omega F(\omega) e^{i \omega t},
\end{equation}
where the domain of integration here and henceforth extends over the whole real axis. For the temporal and spectral amplitude to be related by Fourier transformation, we must assume $\frac{\omega_0}{\Delta \nu} \gg 1$; that is, the spectral amplitude must have (at least approximately) support only for positive frequencies.

A delay $\tau$ in the expected time of arrival of the signal which could be caused by synchronization errors can then be introduced as $t \mapsto t-\tau$.
In the spectral domain, this delay acts as a complex rotation
\begin{equation}
    A_\mathrm{rec}(t)= A(t - \tau) = \frac{1}{\sqrt{2\pi}}\int d\omega F(\omega) e^{i \omega (t - \tau)}.
\end{equation}
The combined effect of Doppler and delay are
\begin{align}
    F_\mathrm{rec}(\omega) &= \sqrt{1+z} F\left((1+z)\omega\right)  e^{-i\omega \tau}
    \label{eq:strech_delay}
    \\
    A_\mathrm{rec}(t) &= A\left(\frac{t-\tau}{1+z}\right) \frac{1}{\sqrt{1+z}} \,.
\end{align}
A stretch in the frequency domain naturally corresponds to a compression in the temporal domain while a shift in the temporal domain corresponds to a complex rotation in the spectral domain.

For realistic communication scenarios, we can expect that the Doppler shift and delay would be estimated and the result would be used to compensate for the effects. However, the estimation and the correction will have a finite accuracy. The residual Doppler shift and delay will consist of a systematic part $\delta$ and a statistical part $\xi$, that is,
\begin{align}
    \label{eq:z_err}
    z_\mathrm{res} &= z - z_\mathrm{est} = \delta_z + \xi_z\\
    \label{eq:tau_err}
    \tau_\mathrm{res} &= \tau - \tau_\mathrm{est} = \delta_\tau + \xi_\tau\,.
\end{align}
The systematic and statistical parts can be associated with the slowly varying and quickly varying contributions, respectively.


\subsection{ Overlap of signal and local reference}\label{sec:overlap}
Whenever a signal is to be coherently compared with a previously agreed-upon reference,
as in the case of CV-QKD,
the spatiotemporal distortions of the signal relative to such a reference are of significance. A measure of correlation that can be used to quantify this distortion is the overlap integral of the signal spectrum with the reference spectrum.
Quantum mechanically this is equivalent to the fidelity of the signal state with respect to the reference state.
For pure states, which we consider here, the spectral overlap integral is recovered.

If a signal and local oscillator are given in terms of their spectral amplitudes $F_S(\omega), F_{L}(\omega)$, their overlap is defined as the inner product
\begin{equation}
    \Braket{F_L,F_S} = \int F^*_L(\omega) F_S(\omega)  d\omega.
\end{equation}

If the signal and local oscillator now differ solely by a Doppler shift and a delay, we make use of equation \eqref{eq:strech_delay} 
and find
\begin{equation}
\begin{aligned}
\label{eq:amb_function}
    Q(z, \tau) & = \sqrt{1+z} \int F^*_L(\omega) F_L\left((1+z)\omega \right)
    e^{-i\omega \tau}  d\omega
    \\&=
    \sqrt{1+z} \int A_L^{*}((1+z)t + \tau) A_L(t) dt.
\end{aligned}
\end{equation}
$Q(z,\tau)$ is a generalized autocorrelation function of the local oscillator mode which is well known in radar literature as the ambiguity function  
\cite{Woodward1954ProbabilityAI, harger1970synthetic, kellyMatchedFilterTheoryHighVelocity1965, eusticeWoodwardAmbiguityFunction2015} and also appears in gravitational wave analysis \cite{creighton2011gravitational, schutz2012gravitational}. The ambiguity function is the output of a matched filter for a given waveform.
Therefore, the problem in sensing is, in some sense, dual to the problem in communication; what is a nuisance in communication might be the signal in sensing.
We note that the ambiguity function is also the characteristic function of the Wigner-Ville distribution of time-frequency analysis \cite{cohenTimefrequencyAnalysis1995, alonsoWignerFunctionsOptics2011a} which is in turn closely related to the Wigner function of quantum mechanics \cite{Wigner_distribution}.
For the remainder of the article, we will retain the radar terminology and call $Q(z,\tau)$ the ambiguity function.

For narrowband signals $\frac{\omega_0}{\Delta \nu} \gg 1$ and small Doppler shifts $z\ll 1$, the effect of the Doppler shift on the signal may be approximated by a constant frequency shift of the carrier frequency $\omega_0$ (peak frequency) $\omega_D = z \omega_0$. In this approximation, the modulation (the shape) remains unaltered.
This reduces the general ambiguity function \eqref{eq:amb_function} to the so-called Woodward ambiguity function\cite{kellyMatchedFilterTheoryHighVelocity1965}
\begin{equation}
\begin{aligned}
\label{eq:woodward_amb}
    \chi(\omega_D, \tau) &= \int F^*_{L}(\omega) F_{L}(\omega + \omega_D)  e^{-i \omega\tau} d\omega
    \\
    &
    = \int  A^*_{L}(t+\tau) A_{L}(t)
    e^{-i\omega_D t} dt.
\end{aligned}
\end{equation}
In the remainder of the article, we assume narrowband signals and small Doppler shifts and use only the Woodward ambiguity function.

We shortly note that for zero Doppler shift the ambiguity function reduces to the first order coherence function while for zero delay it reduces to the spectral auto-correlation function.

\section{Doppler shift and delay in CV-QKD \label{sec:cvqkd}}

\subsection{Homodyne detection}

Here, we illustrate the occurrence of the overlap, i.e. mode match, in quantum communication. 
To this end, we consider continuous-variable quantum key distribution (CV-QKD) in free space between two parties, Alice and Bob who are using a coherent state protocol with a \emph{local} local oscillator (LLO) design \cite{marieSelfcoherentPhaseReference2017a}.
The LLO design was introduced to eliminate the security loophole in the transmitted local oscillator (TLO) design \cite{Bing_Locallocaloscillator_CVQKD_2015, Soh_selfreferenced_cvqkd_llo_2015, Huang_locallocaloscillator_cvqkd_2015}.
However, this introduces the possibility of a mode mismatch between the local oscillator and the signal state as they no longer propagate through the same channel.

\begin{figure}
    \centering
    \includegraphics[width=1\linewidth]{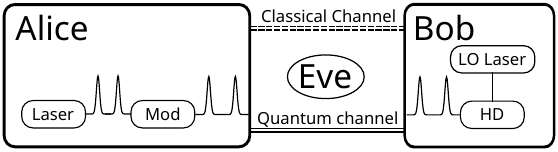}
    \caption{Sketch of the CV-QKD protocol with a local local oscillator, \cite{marieSelfcoherentPhaseReference2017a}.
    Alice sequentially sends weak signal pulses (red) and intense phase reference pulses (blue). Bob performs a homodyne detection (HD) of both signal and reference pulses in his phase reference system using the local oscillator (LO).
    }
    \label{fig:CVQKD_sketch}
\end{figure}

A sketch of the protocol is shown in Fig. \ref{fig:CVQKD_sketch}.
From Alice's laser source, she generates a train of pulses that are modulated to weak signal and high-intensity reference pulses (as a phase reference is necessary for any quantum information protocol in continuous variables, \cite{bartlettReferenceFramesSuperselection2007}).
The weak signal state's quadratures are modulated according to a complex Gaussian distribution by Alice
\cite{grosshans2002continuous, grosshans2003quantum}.
The slightly delayed reference pulses are carrying phase information to Bob, effectively establishing a common measurement basis.
Bob locally creates a train of local oscillator pulses (at the same repetition rate) that are used to detect the incoming signal and reference pulses coherently. According to his local oscillator, he performs a homodyne measurement on a random quadrature. The quadratures themselves are determined according to Alice's phase reference. Having received all signal states Bob then communicates to Alice his choices of quadrature. Alice only keeps the quadrature values matching Bob's choices, ultimately generating the secret key.
For a comprehensive review see e.g. \cite{weedbrook2012gaussian}.
The rate at which the secret key can be generated is determined by the choice of protocol as well as the properties of the optical quantum channel.
Calculations of key rates for CV-QKD protocols can be found for example in \cite{GP, pirandola2017fundamental}. Fundamental bounds in quantum communication were established in \cite{pirandola2017fundamental}. In \cite{pirandolaLimitsSecurityFreespace2021} sources of noise and loss in satellite quantum communication such as thermal noise, pointing errors, diffraction, and atmospheric loss were investigated as well as their effect on key rates. In the following, we will assume the channel to be ideal except for the Doppler and delay-induced mode mismatch.

The signal and local oscillator are assumed to be in a continuous mode coherent state, which was introduced in Sec. \ref{sec:photonic_wavepackets}, Eq. \eqref{eq:cont_coherent} (see also \cite{blowContinuumFieldsQuantum1990, Mandel_Wolf_1995, vanenkQuantumStatePropagating2001})
\begin{align}
    \label{eq:coherent_signal}
    \Ket{\alpha_{S}}_{S} &= \bigotimes_\omega \Ket{\alpha_{S} F_{S}(\omega)},
    \\
    \label{eq:coherent_LO}
    \Ket{\alpha_{L}}_{L} &= \bigotimes_\omega \Ket{\alpha_{L} F_{L}(\omega)},
\end{align}
where $F_{S}(\omega)$ and $F_{L}(\omega)$ are the normalized spectral amplitude functions of the signal and local oscillator state. $\alpha_{S, L}$ is the complex amplitude with modulus $|\alpha_{S,L}|$ and phase $\theta_{S,L}$.

\begin{figure}
    \centering
    \includegraphics[width=1.\linewidth]{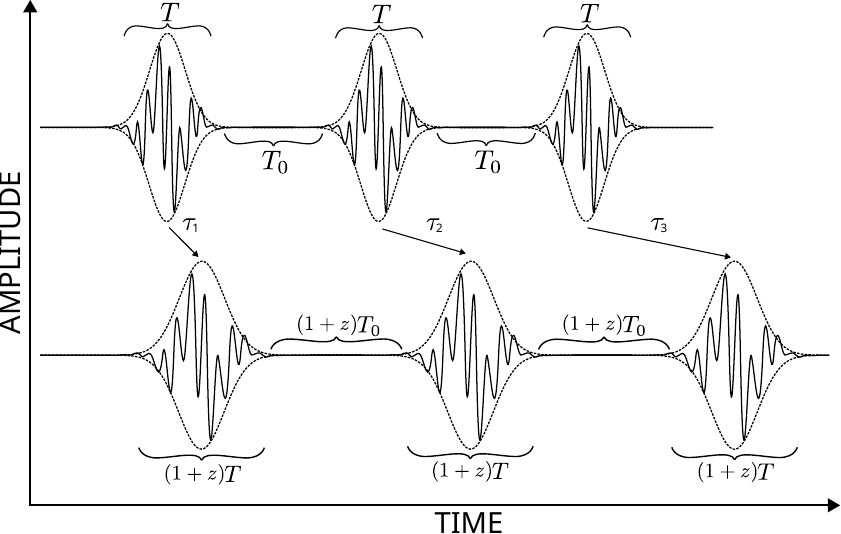}
    \caption{A pulsed signal of length $T$ inside time bins of length $T_0$. Due to Doppler shift the time-bins are dilated by $z$ and acquire an offset $\tau_i$ each contributing to a mode mismatch.}
    \label{fig:time_bins_mismatch}
\end{figure}

We assume the presence of mode mismatch defined as $1-\gamma > 0$ through the generally complex overlap integral
\begin{equation}
\gamma := \int d\omega F_S(\omega) F^*_{L}(\omega)\,,
\end{equation}
also called the mode match.
In Fig. \ref{fig:time_bins_mismatch} we illustrate how a mismatch in the temporal and spectral domain might occur: A signal pulse is sent in each time bin of length $\delta t$. Due to Doppler shift the time bins are dilated such that the length is now $z \delta t$. Due to this dilation (and other synchronization errors), there is an offset $\tau$ of the incoming signals in addition to the Doppler dilation of each pulse.

\begin{figure}
    \includegraphics[width=0.8\linewidth]{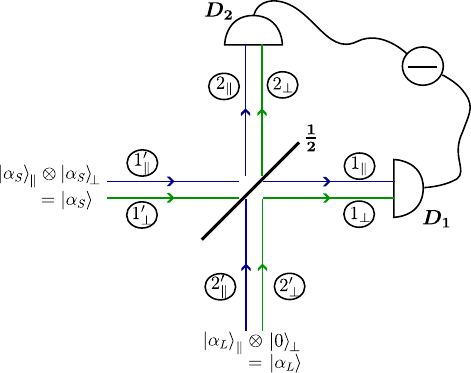}
    \caption{Mode decomposed balanced homodyne detection scheme for the signal state $\Ket{\alpha_{S}}$ and the local oscillator $\Ket{\alpha_{L}}$. Blue and green lines indicate the mode parallel or perpendicular to the local oscillator mode. Circled numbers indicate the mode.}
    \label{fig:detection}
\end{figure}

Upon receiving the signal, Bob performs balanced homodyne detection as depicted in Fig. \ref{fig:detection}.
The signal and Bob's local oscillator are directed at two ports of a balanced beam splitter and the intensities at the output ports are measured.
To simplify the analysis, we decompose the input
states into spectral modes parallel and orthogonal to the local oscillator mode instead of using a decomposition into monochromatic modes.
As in \cite{lordi2023_Quantum_theory_temporally_mismatched_homodyne} we find
\begin{align}
    \Ket{\alpha_{1'}}_{1'} &= \Ket{\gamma \alpha_S}_{1'_\parallel}
    \otimes \Ket{\sqrt{1-|\gamma|^2} \alpha_S}_{1'_\perp},
\\
    \Ket{\alpha_{2'}}_{2'} &= \Ket{\alpha_{L}}_{2'_\parallel} \otimes \Ket{0}_{2'_\perp}.
\end{align}
Coherent states transform in a beamsplitter simply by transforming their amplitude like classical electromagnetic fields \cite{tycOperationalFormulationHomodyne2004}, that is, 
\begin{equation}\label{eq:BS}
 \begin{split}
    \alpha_1 = \frac{\alpha_{1'} + i \alpha_{2'}}{\sqrt{2}} \quad\mathrm{and}\quad    \alpha_2 = \frac{i \alpha_{1'} + \alpha_{2'}}{\sqrt{2}} .
\end{split}
\end{equation}
Therefore, we find for the output states 
\footnote{We assume here that each spectral mode transforms independently in the beamsplitter. This was shown in \cite{lordi2023_Quantum_theory_temporally_mismatched_homodyne}.}
\begin{align}
\label{eq:outputs}
    \Ket{\alpha_1}_1
    &=
    \Ket{\frac{\gamma \alpha_S + i \alpha_{L}}{\sqrt{2}}}_{1_\parallel}
    \otimes 
    \Ket{\frac{\sqrt{1-|\gamma|^2} \alpha_S}{\sqrt{2}}}_{1_\perp},
\\
    \Ket{\alpha_2}_2 
    &=
    \Ket{\frac{i \gamma \alpha_S + \alpha_{L}}{\sqrt{2}}}_{2_\parallel}
    \otimes 
    \Ket{\frac{ i\sqrt{1-|\gamma|^2} \alpha_S}{\sqrt{2}}}_{2_\perp}\,.
\end{align}

Assuming idealized detectors that do not discriminate spectral modes (i.e. have a flat detector response), the statistics of the photocurrents, which are the quantity measured by Bob, are determined by the photon numbers in each output mode.
The total photon number operator is given by the sum of the individual photon number operators for parallel and perpendicular mode \cite{blowContinuumFieldsQuantum1990}
\begin{equation}
    {n_{1,2}} = {n_{{(1,2)}_\parallel}} + {n_{{(1,2)}_\perp}}.
\end{equation}
For the coherent states under consideration, we find
\begin{align}
    \Braket{n_1} &= 
    \frac{1}{2} (
    |\alpha_{L}|^2 + |\alpha_S|^2 
    - i \alpha_{L}^* \alpha_S \gamma
    + i \alpha_{L} \alpha_S^* \gamma^*
    )
    \\
    \Braket{n_2} &= 
    \frac{1}{2} (
    |\alpha_{L}|^2 + |\alpha_S|^2 
    + i \alpha_{L}^* \alpha_S \gamma
    - i \alpha_{L} \alpha_S^* \gamma^*
    ).
\end{align}
The difference in photon numbers is then
\begin{equation}
\begin{aligned}
    \Braket{\Delta n} &= 2 \Im(\gamma^* \alpha_S^* \alpha_{L}),
    \\
    &= 2 |\alpha_S \alpha_{L} \gamma|
    \sin (\theta_{L} - \theta_S - \Gamma)
    \label{eq:exp_diff}
\end{aligned}
\end{equation}
where $\Gamma = \arg \gamma$ is the phase of the overlap integral.
From this, we see that the mode mismatch acts as an effective loss in the amplitude and a phase shift of the signal beam through the modulus and the argument of the overlap $\gamma$, respectively.

To find the signal-to-noise ratio (SNR), we first evaluate the variance $\sigma^2$
\begin{equation}
    \label{eq:variance_diff}
    \sigma^2 = \Braket{(\Delta n)^2} - \Braket{\Delta n}^2
    = |\alpha_S|^2 + |\alpha_{L}|^2,
\end{equation}
which is the expected shot noise due to signal and local oscillator intensity.
The SNR then is
\begin{equation}
    \label{eq:snr}
    \mathrm{SNR} = \frac{|\Braket{\Delta n}|}{\sigma}
     = \frac{2 |\alpha_S \alpha_{L} \gamma \sin(\theta_{L} - \theta_S - \Gamma)|}{\sqrt{|\alpha_{S}|^2 + |\alpha_{L}|^2}},
\end{equation}
which in the limit of a strong local oscillator reduces to
\begin{equation}
    \label{eq:snr2}
    \mathrm{SNR} = 2|\alpha_S \gamma \sin(\theta_{L} -\theta_S-\Gamma)| = 2 |\Im (\gamma \alpha_S)|.
\end{equation}
Except for a phase shift of $\frac{\pi}{2}$ originating from the choice of beamsplitter transformation this is the same result as obtained in \cite{lordi2023_Quantum_theory_temporally_mismatched_homodyne} and is analogous to the results of \cite{shapiroQuantumTheoryOptical2009} for spatial mode mismatch.

\begin{figure}
    \includegraphics[width=0.8\linewidth]{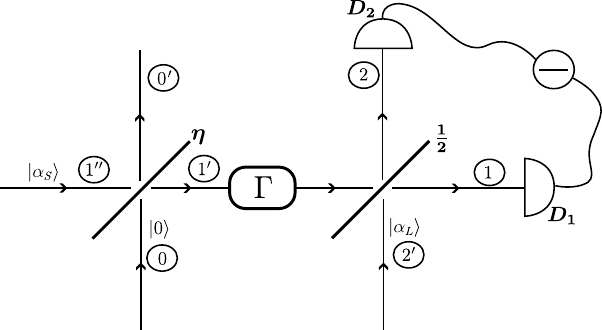}
    \caption{Balanced homodyne detection scheme with an $\eta$-transmissivity beamsplitter and $\Gamma$-phase shifter in front. Circled numbers indicate the mode.}
    \label{fig:detection_loss}
\end{figure}

For a much more detailed treatment of the statistics of difference counts in homodyne detection without the assumption of a strong local oscillator, we refer to \cite{vogelStatisticsDifferenceEvents1993, lipfertHomodyneDetectionOnoff2015, tycOperationalFormulationHomodyne2004}.
For a more detailed treatment of the homodyne detection with a temporal mismatch, we refer to \cite{lordi2023_Quantum_theory_temporally_mismatched_homodyne}.

\subsection{Quantum channel description}
Here we will shortly demonstrate that the mode mismatch in homodyne detection can be shown to be equivalent to an effective loss and a phase shift in the signal channel in the limit of a strong local oscillator.
The corresponding setup for a lossy signal channel with no mode mismatch implemented by a beamsplitter of transmissivity $\eta$ and a $\Gamma$-phase shifter is illustrated in Fig. \ref{fig:detection_loss}.
We first identify the overall states before detection in both setups with each other. In the original setup (fig. \ref{fig:detection}), the overall state has four distinct modes. Rearranging the modes and using \eqref{eq:outputs}, we find 
\begin{align}
    \nonumber \Ket{\phi} = &\Ket{\frac{\gamma \alpha_S + i \alpha_{L}}{\sqrt{2}}}_{1_\parallel}
    \otimes 
    \Ket{\frac{i \gamma \alpha_S + \alpha_{L}}{\sqrt{2}}}_{2_\parallel}
    \\&
    \otimes
    \Ket{\frac{\sqrt{1-|\gamma|^2} \alpha_S}{\sqrt{2}}}_{1_\perp}
    \otimes 
    \Ket{\frac{i \sqrt{1-|\gamma|^2} \alpha_S}{\sqrt{2}}}_{2_\perp}.
\end{align}
For the lossy dephasing setup (Fig. \ref{fig:detection_loss}), the state is
\begin{equation}
\begin{split}
    \nonumber \Ket{\varphi} = \Ket{\frac{\eta e^{i\Gamma} \alpha_S + i \alpha_{L}}{\sqrt{2}}}_{1}
    & \otimes \Ket{\frac{i \eta e^{i\Gamma} \alpha_S + \alpha_{L}}{\sqrt{2}}}_{2}
    \\ &
    \otimes \Ket{\frac{\sqrt{1-\eta^2}\alpha_S}{\sqrt{2}}}_{0'}.
\end{split}
\end{equation}
In these two states, we can identify the perpendicular modes in $\mathcal{H}_{1_\perp} \otimes \mathcal{H}_{2_\perp}$ with the undetected beamsplitter output $\mathcal{H}_{0'}$.
We now assume that these modes are not seen by the detector and trace them out.
Identifying the magnitude of the overlap with the transmissivity $\eta = |\gamma|$ and the phase with its argument $\Gamma = \arg \gamma$ we find that the states are formally equivalent
\begin{equation}
    \begin{aligned}
    \Ket{\phi_{red}} &= \Ket{\varphi_{red}} \\
    & = \Ket{\frac{\eta e^{i\Gamma} \alpha_S + i \alpha_{L}}{\sqrt{2}}}_{1}
    \otimes 
    \Ket{\frac{i \eta e^{i\Gamma} \alpha_S + \alpha_{L}}{\sqrt{2}}}_{2}.
    \end{aligned}
\end{equation}
Calculating the expectation and variance for the difference operator for this state we find
\begin{align}
    \Braket{\Delta n} = 2 \Im (\eta e^{-i\Gamma} \alpha_S^* \alpha_{L})
    \\
    \sigma^2 = |\eta \alpha_S|^2 + |\alpha_{L}|^2.
\end{align}
Which differs only by the noise of the perpendicular signal mode which the detector does not detect. In the strong local oscillator limit, however, the noise is dominated purely by the local oscillator's shot noise and we find the same detection statistics as for the mode-matched case. We have therefore shown that in the strong oscillator limit, homodyning of a mode mismatched signal is equivalent to the homodyning of a signal that went through a quantum channel applying loss and a phase shift. 
The quantum channel that describes Doppler shift and delay is therefore a lossy dephasing channel, that is, a composition, $\mathcal{L}_\eta \circ \mathcal{N}_p$, of a lossy channel $\mathcal{L}_\eta$ with transmissivity $\eta$ and corresponding loss $1-\eta$
\begin{equation}
    \label{eq:loss_channel}
    \mathcal{L}_\eta: \Ket{\alpha} \mapsto \Ket{\eta \alpha},
\end{equation}
and a dephasing channel $\mathcal{N}_p$
\begin{equation}
    \label{eq:deph_channel}
    \mathcal{N}_p: \Ket{\alpha} \mapsto \int_{-\pi}^\pi p(\Gamma) \Ket{e^{i \Gamma} \alpha} d\Gamma,
\end{equation}
where $p$ is the probability distribution governing the dephasing, \cite{lamiExactSolutionQuantum2023}.

\subsection{Quantum channel capacity}
\label{sec:quantumchannelcapacity}

In Section \ref{sec:deformation}, we have argued that the residual redshift and delay after estimation and compensation can be separated into systematic (low frequency) contributions $\delta_z$ and $\delta_\tau$ and stochastic (high frequency, vanishing mean) contributions $\xi_z$  and $\xi_\tau$. From the result of the last section, we can conclude that these contributions will induce a systematic part of loss 
and phase $\delta_{1-\eta}$ and $\delta_\Gamma$, respectively, and a stochastic part of loss and phase $\xi_{1-\eta}$ and $\xi_\Gamma$, respectively.
As the systematic phase corresponds to a rotation of the measurement basis (a unitary operation), we assume that this can be compensated in post-processing through appropriate data analysis, such as precise orbit determination \cite{gao2015analysis, Kuchynka_Martin_Serrano_Merz_Siminski_2020, dequal2021feasibility}.
In contrast, the systematic loss $\delta_{1-\eta}$ 
as well as the stochastic contributions cannot be compensated.

Although, to the knowledge of the authors, the private or quantum capacity of the lossy dephasing channel is not yet known, one can find an upper bound for the capacity from the capacities of its substituents, the lossy channel and the dephasing channel; the capacity is always bounded by the lowest capacity of its substituents \cite{lamiExactSolutionQuantum2023}.
The private and quantum capacity of the lossy channel is given by the so-called PLOB bound \cite{pirandola2017fundamental}
\begin{equation}
    \label{eq:cap_plob}
    P(\mathcal{L}_\eta) = Q(\mathcal{L}_\eta) =  -\log_2 (1-\eta).
\end{equation}
The private and quantum capacities of the dephasing channel have recently been found to be given by the relative entropy of the phase distribution to the uniform distribution \cite{lamiExactSolutionQuantum2023}
\begin{equation}
    \label{eq:cap_deph}
    \begin{aligned}
    P(\mathcal{N}_p) & =    Q(\mathcal{N}_p) \\
    & = D(p||u) := \int_{-\pi}^{\pi} d\phi \,p(\phi) \log_2 \frac{p(\phi)}{1/(2\pi)},
    \end{aligned}
\end{equation}
where $p(\phi)$ is the probability distribution for the phase such that it is wrapped to the interval $[-\pi, \pi]$.
The capacities of the lossy dephasing channel are then upper bounded by
\begin{align}
    \label{eq:loss_deph_capacity}
    \nonumber Q(\mathcal{L}_\eta \circ \mathcal{N}_p) &\leq P (\mathcal{L}_\eta \circ \mathcal{N}_p)\\
    &
    \leq \min \big\{ -\log_2 (1-\eta), D(p||u)\big\}.
\end{align}

\subsubsection{Systematic loss}
The systematic loss $\delta_{1-\eta}$ corresponds to the contribution of systematic mode mismatch originating from the systematic errors in timing (delay) $\delta_\tau$ and frequency (Doppler) $\delta_z$, see eq. \eqref{eq:z_err},\eqref{eq:tau_err}. It is given through the magnitude of the ambiguity function for the given systematic delay and Doppler shift values
\begin{equation}
    \delta_{1-\eta} = 1-\left | \chi(\delta_{\omega_D}, \delta_\tau) \right|,
\end{equation} where $\delta_{\omega_D} = \omega_0 \delta_z$.
As already mentioned, we assume that the systematic phase error is known (e.g. measured by pilot pulses) and corrected for in post-processing. Therefore, the phase of the ambiguity function does not contribute to systematic errors.

\subsubsection{Stochastic loss and phase fluctuations}
The aforementioned stochastic errors in the phase correspond to phase fluctuations in the quantum channel that are induced by the stochastic errors in delay $\xi_\tau$ and Doppler $\xi_z$, see eq. \eqref{eq:z_err}, \eqref{eq:tau_err}. The probability distributions of phase and transmissivity are $p_\phi$ and $p_\eta$, respectively. Due to these fluctuations, the channel becomes a composition of dephasing and fading lossy channels. The capacity of the pure-loss fading channel $P(\mathcal{L}_{p_\eta})$ is given by \cite{pirandolaLimitsSecurityFreespace2021}
\begin{equation}
    \label{eq:fading_loss_plob}
    P(\mathcal{L}_{p_\eta}) = -\int d\eta \, p_\eta(\eta) \log_2 (1-\eta).
\end{equation}
The capacity of the composite channel is then, as before, upper bounded by the minimum of either channel's capacity
\begin{equation}
    \label{eq:fading_loss_deph_capacity}
    P (\mathcal{L}_{p_\eta} \circ \mathcal{N}_{p_\phi})
    \leq \min \big\{ P(\mathcal{L}_{p_\eta})
    , P(\mathcal{N}_{p_\phi})\big\}.
\end{equation}

\section{Pulse shape optimization} 
\label{sec:pulse_shape_optimization}

The actual spectral distribution $F(\omega)$ encountered in experiments of course depends on the source of the radiation as well as many other experimental parameters.
In the following, we will consider the most commonly encountered shapes, Gaussian and Lorentzian functions.
The Lorentzian is divided into single- and double-sided Lorentzian functions.
The single-sided Lorentzian corresponds to the (spontaneous) emission process of excited atoms which decay with a one-sided exponential probability and therefore appear in e.g. quantum dots as well as the transmission function of an optical cavity \cite{santoriIndistinguishablePhotonsSinglephoton2002}, \cite[ch. 2.2]{herskindCavityQuantumElectrodynamics}.
The double-sided Lorentzian occurs for example in the cavity-enhanced spontaneous down conversion process (SPDC) \cite{ouCavityEnhancedSpontaneous1999, collettSqueezingIntracavityTravelingwave1984, collett_quantum_1987, mitchellParametricDownconversionWaveequation2009} commonly used in quantum information processing applications.
Gaussian functions are commonly used to approximate the temporal shape of pulses originating from actively mode-locked lasers \cite{Paschottagaussian_pulses}. Via the Fourier transformation, the Gaussian temporal shape corresponds to a Gaussian spectral amplitude function and spectrum.
For quantum information processing purposes, a single-photon source with a specific lineshape can be realized from the SPDC process by applying a filter on one photon which projects the heralded photon onto the desired spectrum \cite{rohde2005optimal, ouParametricDownconversionCoherent1997a, kolchinElectroOpticModulationSingle2008,peerTemporalShapingEntangled2005, buckleyEngineeredQuantumDot2012}.
In addition to pair sources based on SPDC, single photons with controllable waveforms and timings can be generated using different atomic systems \cite{kellerContinuousGenerationSingle2004, farreraGenerationSinglePhotons2016}.

\begin{figure}[t]
    \centering
    \includegraphics[width=.8\linewidth]{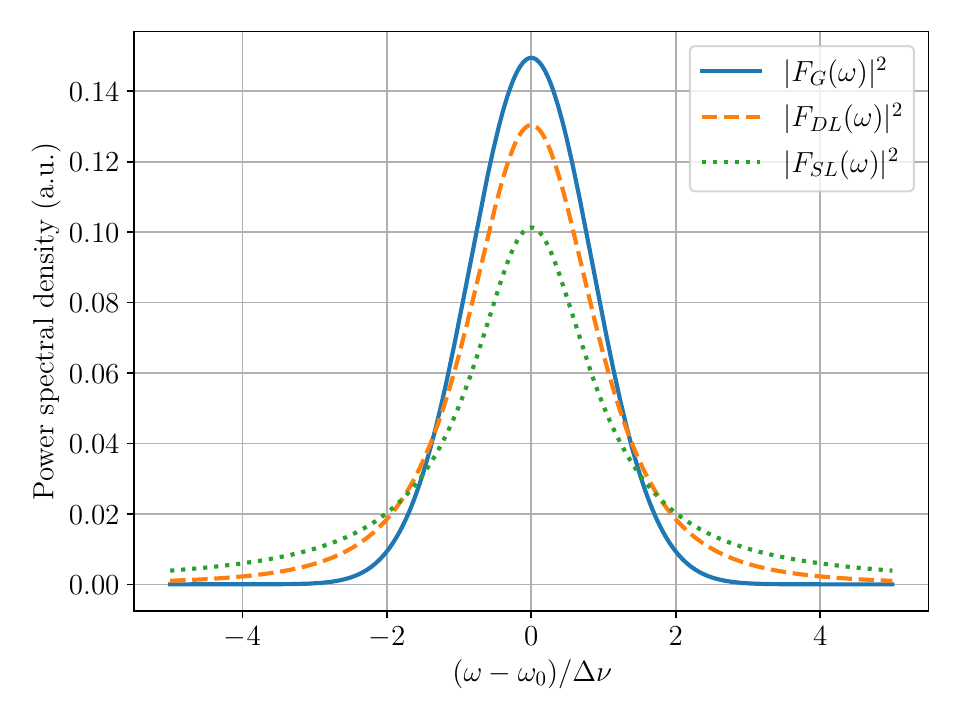}
    \caption{Square of the modulus of the spectral distributions (Power spectral density) given in \eqref{EQ:photmodeFreqProfG}-\eqref{EQ:photmodeFreqProfL}. The frequency is given in units of bandwidth.
    }
    \label{fig:spectra}
\end{figure}

\begin{figure}[t]
    \centering
    \includegraphics[width=.8\linewidth]{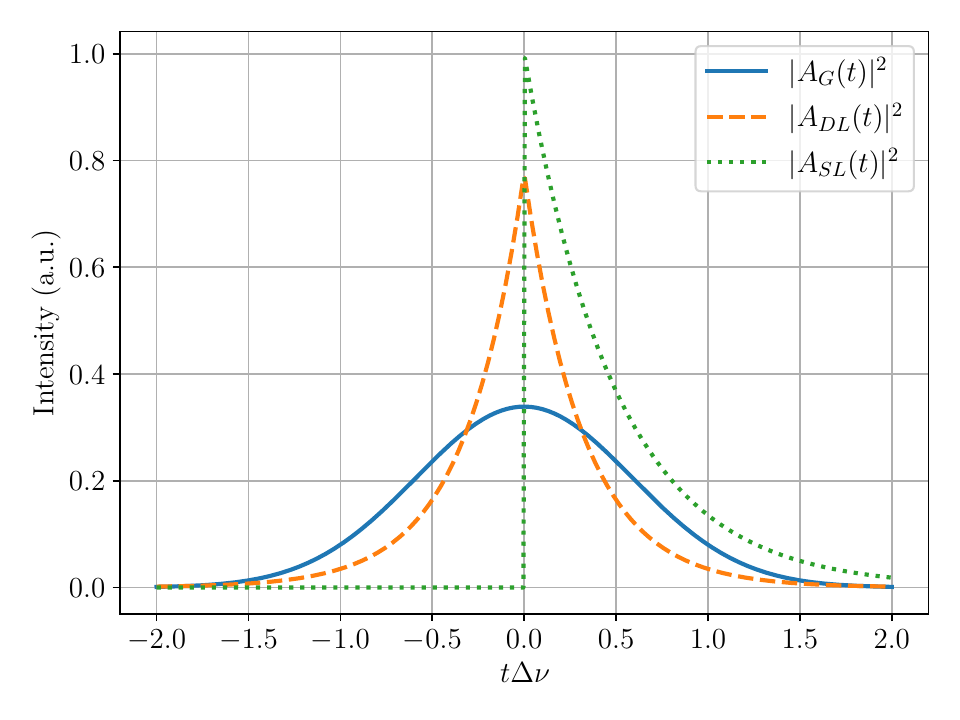}
    \caption{
    Normalized pulse shape in the temporal domain corresponding to the spectral amplitudes \eqref{EQ:photmodeFreqProfG}-\eqref{EQ:photmodeFreqProfL}. The Gaussian spectral amplitude results in a Gaussian pulse shape while the Lorentzians correspond to an exponential shape (single- and double-sided).
    The pulses are centered at $t=0$ and time is given in units of inverse bandwidth.
    }
    \label{fig:signals}
\end{figure}

The normalized spectral amplitudes are given by
\begin{align}
\text{Gaussian: }&\quad F_{G}(\omega)=\frac{1}{\sqrt[4]{2\pi\sigma^2}}e^{-\frac{(\omega-\omega_0)^2}{4\sigma^2}},
\label{EQ:photmodeFreqProfG}
\\
\begin{tabular}{r}
     Lorentzian
     \\
     single-sided:
\end{tabular}
& \quad F_{SL}(\omega)=\sqrt{\frac{\Delta\nu}{\pi}}\frac{1}{\Delta\nu+i(\omega-\omega_0)},
\label{EQ:photmodeFreqProfLI}
\\
\begin{tabular}{r}
     Lorentzian
     \\
     double-sided:
\end{tabular}
& \quad F_{DL}(\omega)= \sqrt{\frac{2s}{\pi}}\frac{s}{s^2+(\omega-\omega_0)^2},
\label{EQ:photmodeFreqProfL}
\end{align}
where $\omega_0$ is the peak-frequency, and $\sigma$, $s$ and $\Delta\nu$ are width parameters of the distributions. The associated signals in the temporal domain are depicted in Fig. \ref{fig:signals}.
The spectrum is the absolute square of the spectral amplitudes, shown in Fig. \ref{fig:spectra}.
To make the distributions comparable we choose the parameters such that all spectra have equal bandwidth, which we define as the Half Width at Half Maximum (HWHM) $\Delta\nu$.
For the Gaussian distribution, the standard deviation is related to the HWHM by $\Delta \nu = \sigma \sqrt{\ln 4}$.
For the double-sided Lorentzian it is $\Delta \nu = s \sqrt{\sqrt{2}-1}$.


\subsection{
Ambiguity Function }
\label{sec:amb_function}

As discussed in Sec.  \ref{sec:cvqkd}, we interpret the absolute value of the ambiguity function as an effective transmissivity of the quantum channel and its complex argument as a phase shift, a misalignment of the measurement bases.
In this section, we give expressions and plots for the ambiguity functions for the three different spectral amplitudes and discuss them.
We consider the general case of combined Doppler shift and delays and the limiting cases of pure Doppler shifts and pure delays.
We restrict our considerations to the 
regime of narrow bandwidth and small redshift and the Woodward ambiguity function in Eq. \eqref{eq:woodward_amb}.
In that case, we can analytically calculate the ambiguity function for the spectral amplitudes of interest (Gaussian and single- and double-sided Lorentzians). Here we only state the results. The derivations of the ambiguity functions \eqref{eq:amb_function} can be found in the App. \ref{app:amb_fun}.

\subsubsection{Combined Doppler shifts and delays}

We start with the general case where both Doppler shifts and delays are present.
For the Gaussian we find
\begin{equation}
    \label{eq:wamb_gauss}
    \chi_{G}(\omega_D, \tau) = e^{-\frac{\omega_D^2}{8 \sigma ^2}-\frac{1}{2} \left(\sigma ^2 \tau^2\right)} 
    e^{- i \tau \left(\omega_0 - \frac{\omega_D}{2}\right)
    }.
\end{equation}
where $\omega_D \equiv \omega_0 z$ is the carrier frequency shift.
For the double-sided Lorentzian, we obtain
\begin{equation}
\begin{aligned}
    \label{eq:wamb_dl}
    \chi_{DL}(z,\tau)& = e^{-s |\tau| 
    -i\tau \left( \omega_0 -\frac{1}{2}\omega_D \right) }
    \frac{
    \cos\left(\frac{\omega_D |\tau|}{2}\right) + \frac{2s}{\omega_D} \sin\left(\frac{\omega_D |\tau|}{2}\right)
    }{1+\left(\frac{\omega_D}{2 s}\right)^2}.
\end{aligned}
\end{equation}
and the ambiguity function for the single-sided 
Lorentzian is given by
\begin{align}
    \label{eq:wamb_sl}
    \chi_{SL}(\omega_D, \tau) =
    \frac{e^{-|\tau| \Delta \nu}}{1 + i\frac{\omega_D}{2\Delta\nu}}
    \begin{cases}
    {e^{-i\tau (\omega_0-\omega_D)}}
    , \tau <0
    \\
    {e^{-i\tau \omega_0}}
    , \tau \geq 0.
    \end{cases}
\end{align}
%
\begin{figure*}
    \centering
    \includegraphics[width=.9\linewidth]{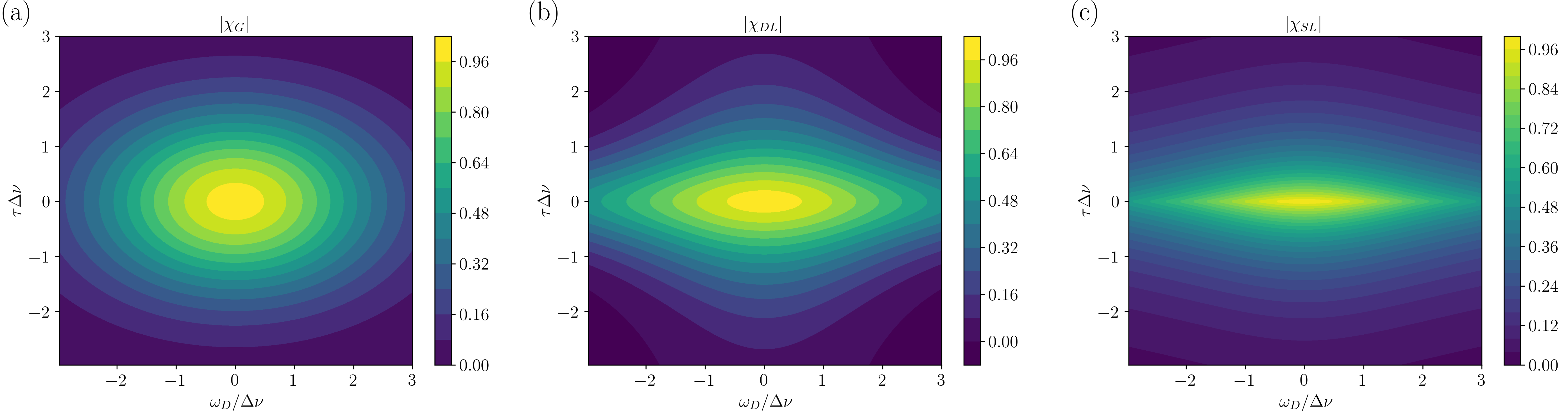}
    \caption{Absolute value of the ambiguity function (mode match, directly related to channel capacities, see Sec. \ref{sec:quantumchannelcapacity}) for Gaussian, double-sided Lorentzian, and single-sided Lorentzian spectra (see Eqns. \eqref{eq:wamb_gauss}, \eqref{eq:wamb_dl}, \eqref{eq:wamb_sl}).
    The Gaussian's ambiguity function has circular contours (or elliptical, depending on the scaling of the axes) indicating that the Gaussian reacts similarly to delay as to Doppler shifts. This is expected as the Gaussian stays a Gaussian under Fourier transformation.
    The double-sided Lorentzians are stretched along the Doppler axes compared to the Gaussian, indicating stronger resilience to Doppler shifts but less against delay. The contours are also concave showing an increased vulnerability to combined Doppler shifts and delays.
    The single-sided Lorentzian is even more robust against Doppler shift in comparison to the Gaussian spectral profile at the cost of an increased vulnerability to delay.
    }
    \label{fig:amb_abs}
\end{figure*}
%
\begin{figure*}
    \centering
    \includegraphics[width=.9\textwidth]{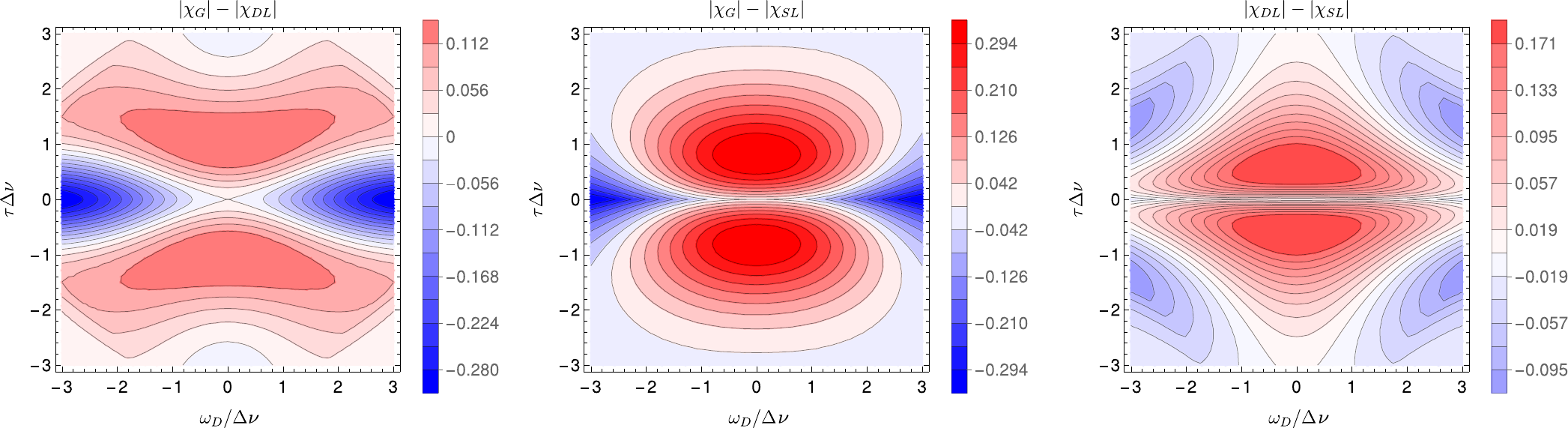}
    \caption{
    Differences in the absolute values of the ambiguity function (mode match, directly related to channel capacities, see Sec. \ref{sec:quantumchannelcapacity}) for various spectra: (a) Gaussian - double-sided Lorentzian (b) Gaussian - single-sided Lorentzian and (c) double-sided Lorentzian - single-sided Lorentzian.
    (a) In comparison to the double-sided Lorentzian, the Gaussian profile achieves higher 
    absolute values of the ambiguity function in the presence of delays while $|\chi_G|$ is smaller 
    for the presence of a pure Doppler shift.
    (b) Qualitatively similar to (a), yet the difference along the delay axis is more significant.
    (c) The double-sided Lorentzian achieves higher  absolute values of the ambiguity function
    along delay axes, while
    $|\chi_{DL}|$ and $|\chi_{SL}|$ do not differ significantly w.r.t. Doppler shifts.
    }
    \label{fig:amb_abs_diff}
\end{figure*}

For the Gaussian and double-sided Lorentzian, the modulus and argument of the ambiguity function can be easily read off from 
 Eqns. \eqref{eq:wamb_gauss}, \eqref{eq:wamb_dl}.
For the single-sided Lorentzian, we find
\begin{align}
    \label{eq:sl_mod}
    |\chi_{SL}(\omega_D, \tau)| &= \frac{e^{-|\tau|\Delta\nu}}{\sqrt{1+(\frac{\omega_D}{2\Delta\nu})^2}},
    \\
    \label{eq:sl_phase}
    \arg \chi_{SL}(\omega_D, \tau) &= 
    \begin{cases}
    -\tau (\omega_0 -\omega_D) - \arctan \left(\frac{\omega_D}{2\Delta\nu}\right)&, \tau<0
    \\
    -\tau \omega_0  - \arctan \left(\frac{\omega_D}{2\Delta \nu}\right)&, \tau \geq0.
    \end{cases}
\end{align}
Fig. \ref{fig:amb_abs} shows the absolute value of the ambiguity function obtained by the different spectral amplitude functions.
Fig. \ref{fig:amb_abs_diff} shows the differences in the absolute values of the ambiguity functions $|\chi|$. In general, we find that they are more sensitive to delays for the Lorentzian spectral profiles than for the Gaussian profile and vice versa with respect to Doppler shifts. A more detailed discussion is given in the figure captions and in the following for the cases of pure Doppler shift and pure delay.

\subsubsection{Pure Doppler shifts}

\begin{figure*}
    \centering
    \includegraphics[width=.9\linewidth]{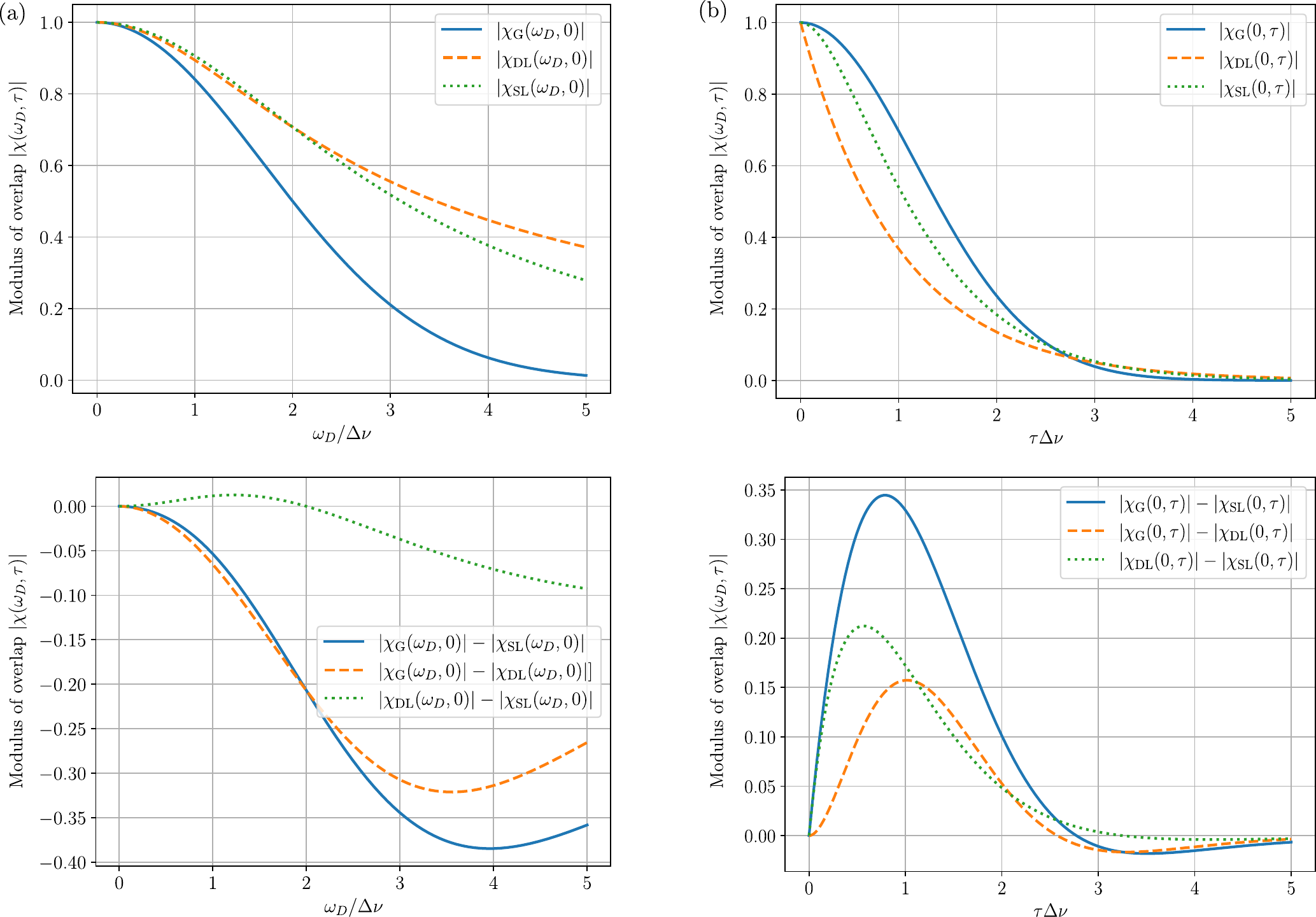}
    \caption{
    Absolute values of the ambiguity functions (top, mode match, directly related to channel capacities, see Sec. \ref{sec:quantumchannelcapacity}) and their differences (bottom) as functions of (a) varying Doppler shift at vanishing delay and (b) varying delay at vanishing Doppler shift
    for three different spectral profiles: Gaussian (blue), double-sided Lorentzian (orange), and single-sided Lorentzian (green)
    (see Eqns. \eqref{eq:wamb_gauss}, \eqref{eq:wamb_sl}, \eqref{eq:wamb_dl})
    (a) It is clearly visible that the Gaussian spectral profile leads to a stronger vulnerability to Doppler shifts than the Lorentzian profiles which perform very similarly.
    (b) We find a superior performance in terms of mode match of the Gaussian spectral profile in comparison to the Lorentzian profiles for small values of $\tau$. This turns into the opposite for larger delays $\tau/\Delta\nu \gtrsim 2.5 $, where the absolute value of the ambiguity function is already quite small in general, however.
}
    \label{fig:overlaps}
\end{figure*}

Here 
we examine the scenario involving only the Doppler shift and assume no delay, that is, a perfect synchronization but imperfect spectral mode match. The ambiguity functions are obtained from those above by setting $\tau=0$. Plots are shown in 
Fig. \ref{fig:overlaps}a.
%
%
%
%
%
Regarding the absolute value of the 
ambiguity function, the Gaussian profile performs worst while the Lorentzians perform quite similarly. 
At small Doppler shifts $|\chi_{DL}(\omega_D,0)|$ for the double-sided Lorentzian lies above $|\chi_{SL}(\omega_D,0)|$ for the singled-sided Lorentzian while for larger Doppler shifts $|\chi_{SL}(\omega_D,0)|$ is the largest.
In regions of significant Doppler shift, say $\omega_D \approx 5\Delta\nu$, $|\chi_G|$ almost vanishes while the single- and double-sided Lorentzians exhibit $|\chi_{SL}|\sim 38\,\%$ and  $|\chi_{DL}|\sim 30\,\%$, respectively.
These results indicate that, in scenarios where the Doppler shift is most relevant, signals with a Lorentzian spectral profile are strongly preferable to a Gaussian one if the absolute value of the ambiguity function and the corresponding mode mismatch are the limiting factors for the performance of the quantum channel. 

\subsubsection{Pure delays}
For the special case of pure delay $\tau$ and no spectral deformation, corresponding to the case of a timing error, we reproduce in essence the results of \cite{rohde2005optimal}.
The temporal overlap $|\chi|$ as a function of delay $\tau$ for the different spectra is shown in 
Fig. \ref{fig:overlaps}b.
The quantitative differences between the profiles in terms of $|\chi|$ are shown in Fig. \ref{fig:overlaps}b (bottom).
It is evident that the Gaussian spectral profile provides the highest absolute value of the ambiguity function (corresponding to the smallest effective loss) in the presence of delays that are smaller than about $\tau \lesssim 2.6/\Delta\nu$.
For larger delays $\tau \gtrsim 3/\Delta\nu$, however, both Lorentzian profiles achieve a higher absolute value of the ambiguity function,
while they do not differ significantly among themselves.

\subsection{Quantum Channel capacities} 
\label{sec:quantum_capacities}

Based on the above expressions for the overlap integrals for the three different spectral functions, we are now ready to investigate the bounds of the quantum channel capacity and analyze the possibilities for their optimization. First, we will investigate the general case of combined Doppler shift and delay, and later, consider the cases of pure Doppler shift and pure delay in separate sub-sections.

\subsubsection{Combined stochastic Doppler shift and delay} 

\begin{figure*}[ht]
    \centering
    \begin{minipage}[b]{0.32\linewidth}
        \centering
        \includegraphics[width=1.\linewidth]{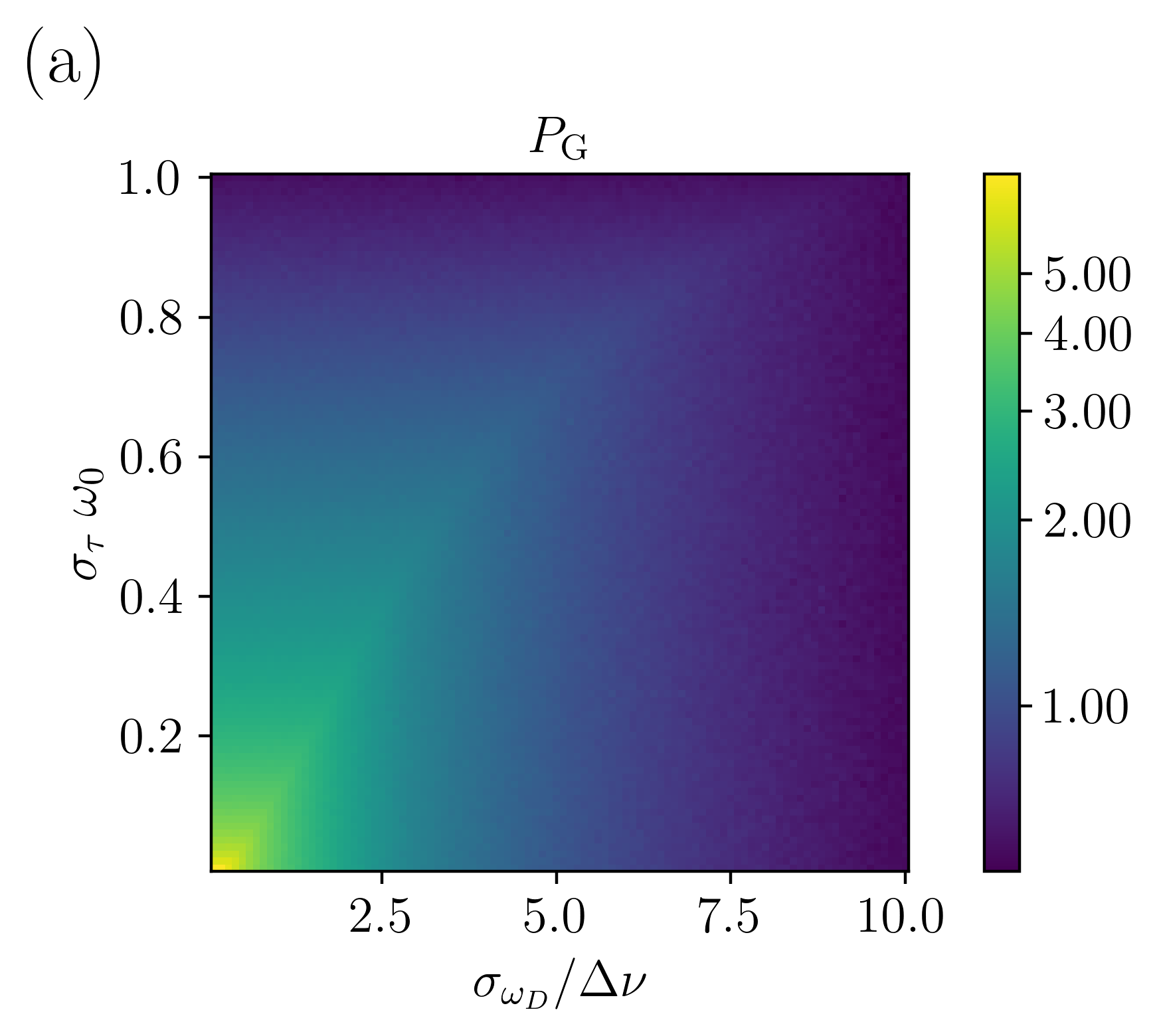}
    \end{minipage}
    \hfill
    \begin{minipage}[b]{0.32\linewidth}
        \centering
        \includegraphics[width=1.\linewidth]{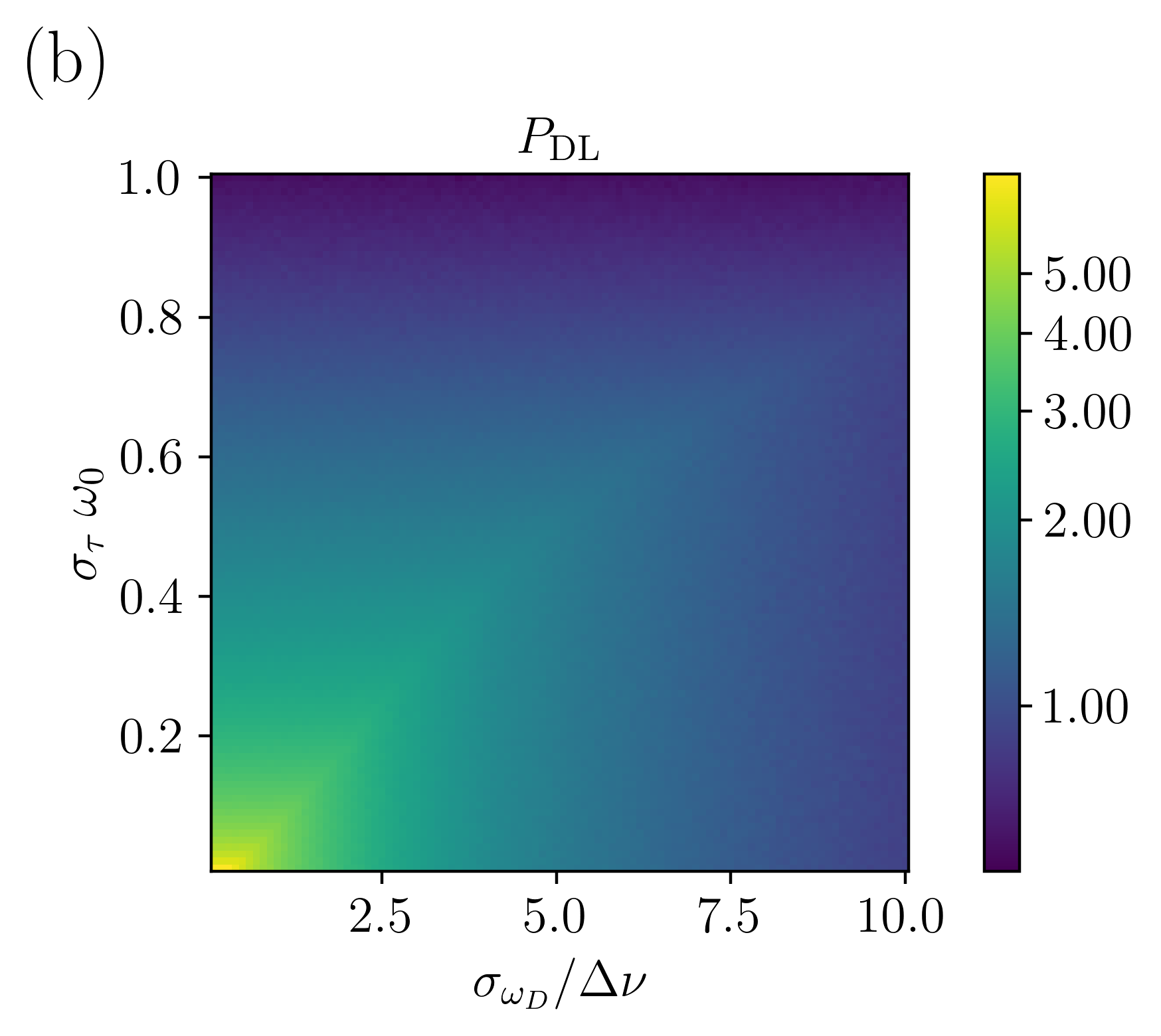}
    \end{minipage}
    \hfill
    \begin{minipage}[b]{0.32\linewidth}
        \centering
        \includegraphics[width=1.\linewidth]{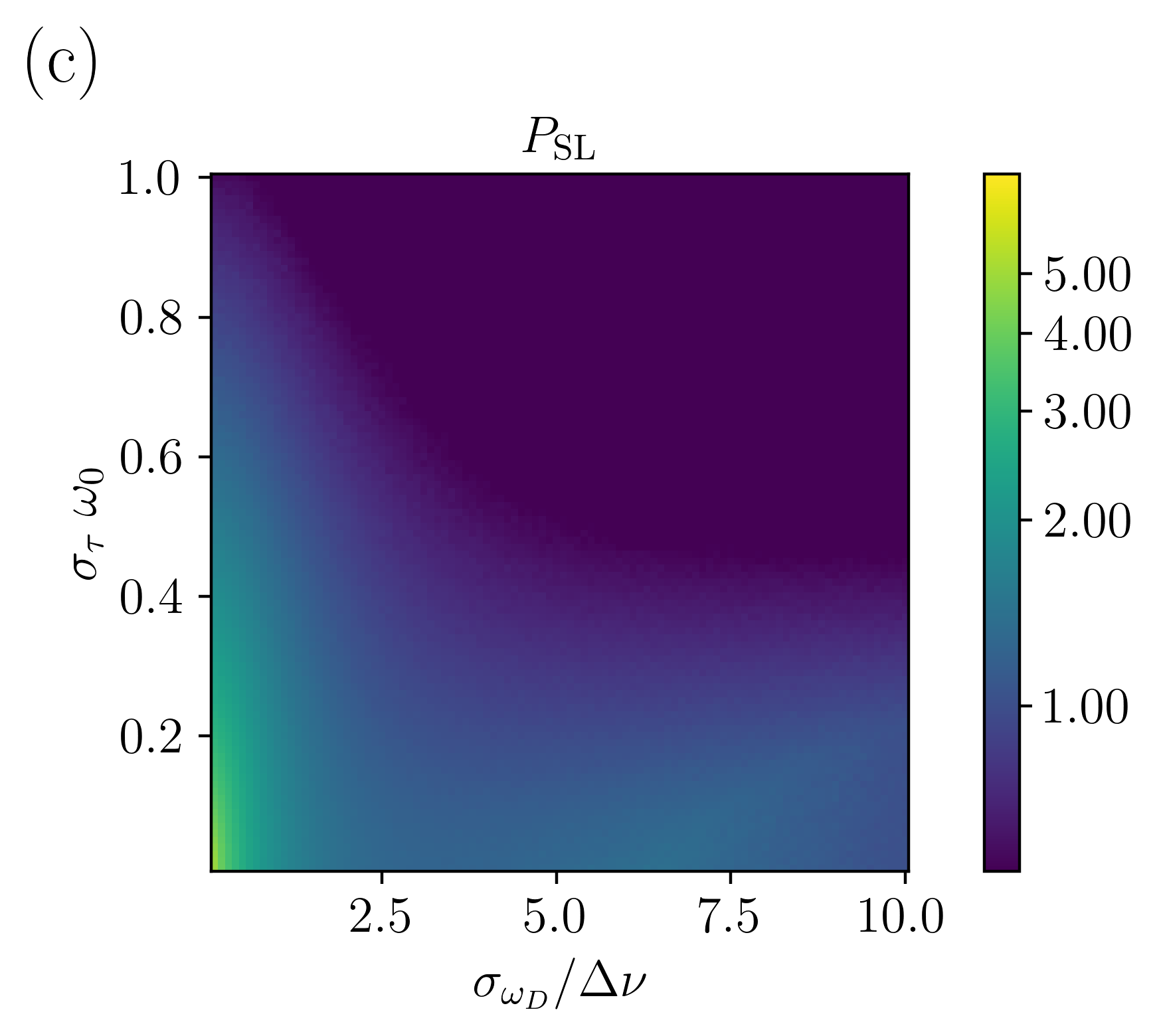}
    \end{minipage}
    \caption{Private capacity bounds (see Eq. \eqref{eq:fading_loss_deph_capacity}) for varying variances of fluctuations in delay and Doppler shift. Each figure displays the capacity bounds corresponding to the spectral profiles
    (a) Gaussian, (b) double-sided Lorentzian, (c) single-sided Lorentzian.
    The Gaussian's and double-sided Lorentzian's behave qualitatively similar. The single-sided Lorentzian's capacity bound shows a sharp decay for Doppler fluctuations since it picks up additional phase fluctuations for Doppler shifts, as explained in Sec. \ref{sec:amb_function}.
    }
    \label{fig:fluc_capacity}
\end{figure*}

Here we investigate the combined effect of both delay and Doppler shift on the quantum capacity of the fading lossy dephasing channel. To treat phase fluctuations and Doppler shifts on the same footing, we consider only fluctuations, i.e. only the stochastic contributions. For this, we assume that delay and Doppler shift originate from normal distributions with zero mean $\xi_{\omega_D} \in \mathcal{N}(0, \sigma_{\omega_D})$, $\xi_{\tau} \in \mathcal{N}(0,\sigma_\tau)$.
We calculated the ambiguity function of each spectral profile from the values of delay and Doppler shift, and subsequently, we obtained the upper bounds for the channel capacity according to Eq. \eqref{eq:fading_loss_deph_capacity}.

In Fig. \ref{fig:fluc_capacity}, the upper bounds for the capacities are depicted.
\footnote{The values of the capacity bounds were calculated numerically by drawing $N=10^6$ samples of delays and Doppler shifts from normal distributions of given variances, each centered at zero.}
For the Gaussian profile and the double-sided Lorentzian profile, the density plots are split into an upper left region with approximate homogeneity along the (horizontal) Doppler shift axis and a lower right region with approximate homogeneity along the (vertical) delay axis.
This is expected since the phase fluctuations are independent of the Doppler shift while the transmissivity remains approximately constant
since the considered delay fluctuations are small compared to the pulse length.
The capacity bound \eqref{eq:fading_loss_deph_capacity} is given by the minimum of either the lossy channel's or the dephasing channels' capacity. The diagonal boundary between the two regions corresponds to the parameter values at which both values coincide.
The capacity bound for the single-sided Lorentzian displays a more complex structure. This is due to the dependence of the phase on the Doppler shift, as seen in Eq. \eqref{eq:sl_phase}.

\begin{figure*}[ht]
    \centering
    \begin{minipage}[b]{0.32\linewidth}
        \centering
        \includegraphics[width=1.\linewidth]{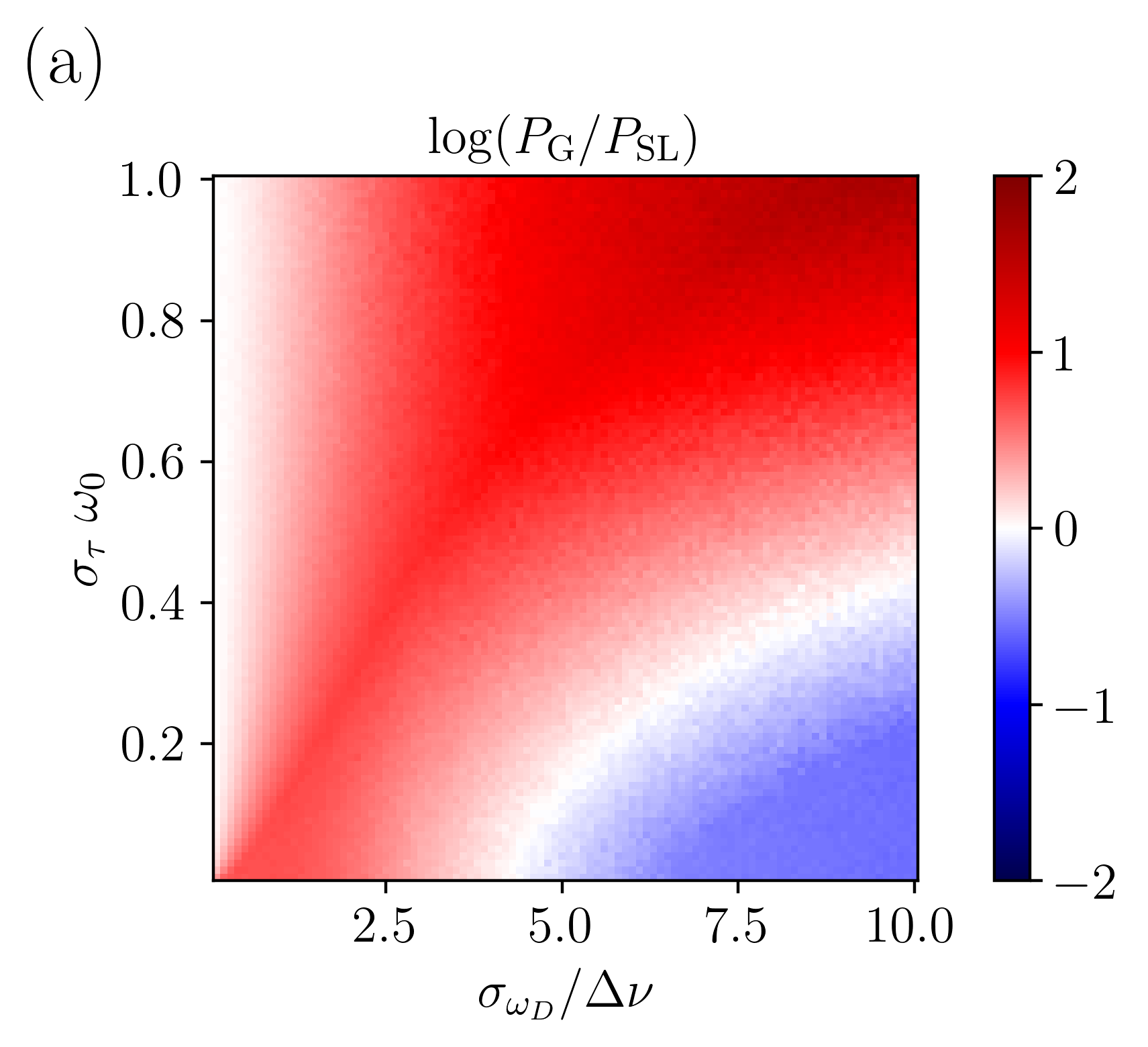}
        \label{fig:subfig1}
    \end{minipage}
    \hfill
    \begin{minipage}[b]{0.32\linewidth}
        \centering
        \includegraphics[width=1.\linewidth]{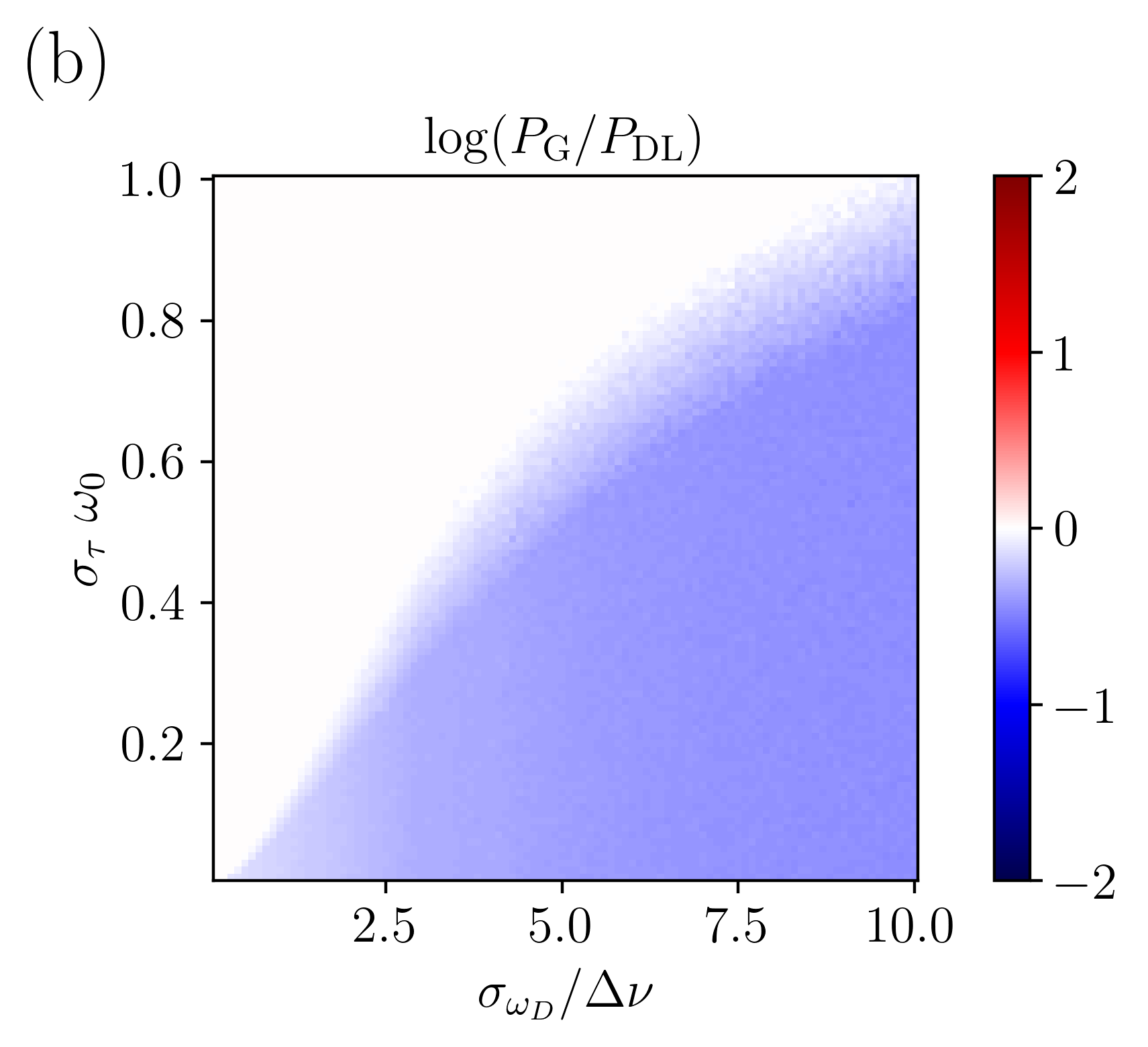}
        \label{fig:subfig2}
    \end{minipage}
    \hfill
    \begin{minipage}[b]{0.32\linewidth}
        \centering
        \includegraphics[width=1.\linewidth]{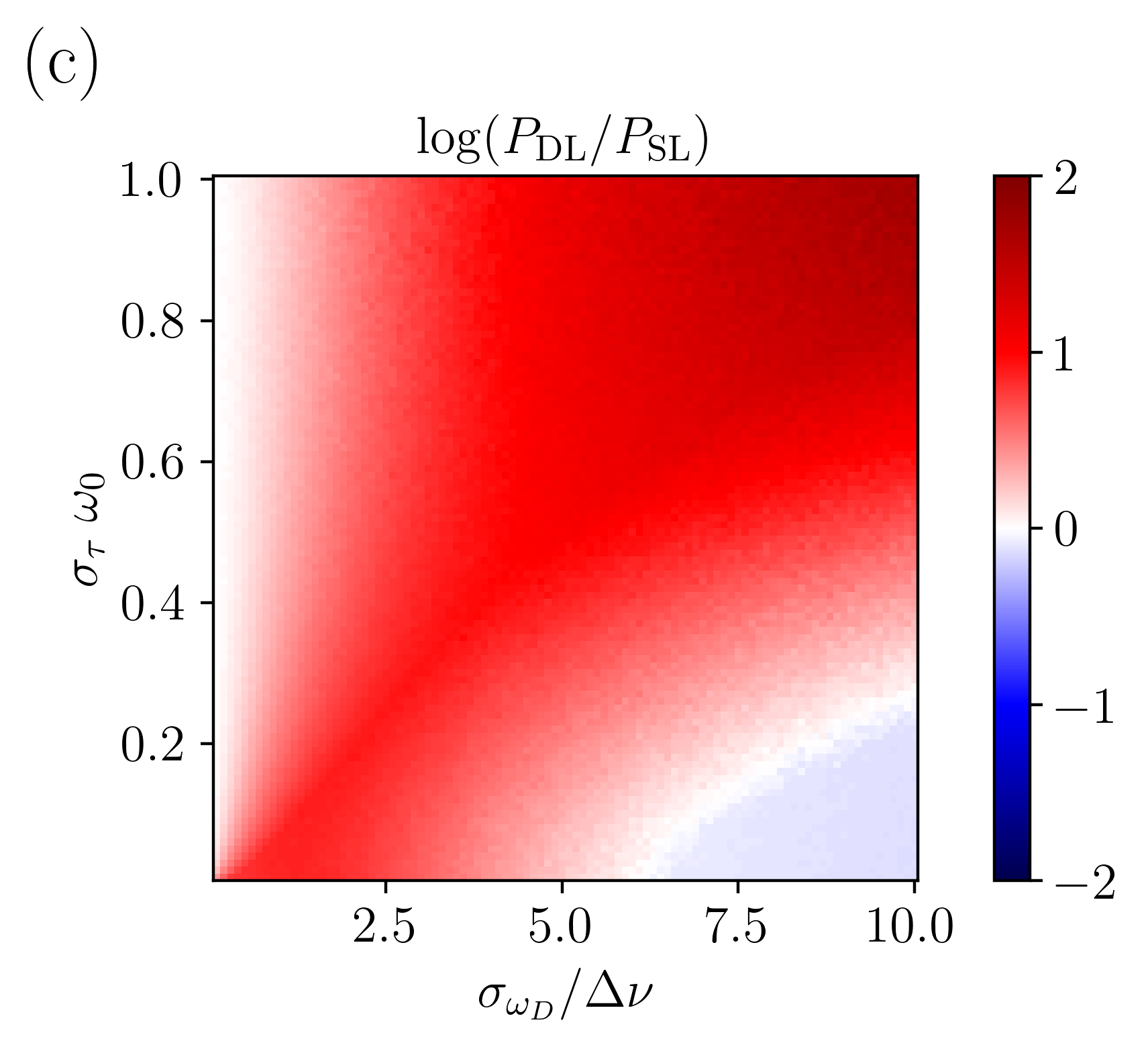}
        \label{fig:subfig3}
    \end{minipage}
    \caption{
    Logarithm of the quotient (equivalent to the difference of logarithms) of the private capacity bounds, see Eq. \eqref{eq:fading_loss_deph_capacity}, between the different profiles:
    (a) $\log(P_{G}/P_{SL})$, (b) $\log(P_{G}/P_{DL})$, (c) $\log(P_{DL}/P_{SL})$. (a) 
    The Gaussian profile leads to higher capacity bounds in regions of small to intermediate Doppler fluctuations than the single-sided Lorentzian profile.
    It also shows higher robustness against fluctuations in delay.
    (b)
    The double-sided Lorentzian shows higher capacity bounds than the Gaussian if Doppler fluctuations are present.
    In terms of robustness against delay fluctuations, they do not differ significantly.
    (c) 
    The double-sided Lorentzian provides higher capacity bounds (similar to the Gaussian in (a)) except for substantial Doppler fluctuations at small delay fluctuations.
    }
    \label{fig:fluc_capacity_diff}
\end{figure*}

To highlight the difference between the profiles, in Fig. \ref{fig:fluc_capacity_diff} we show the differences in the capacity bounds for the profiles.
In Fig. \ref{fig:fluc_capacity_diff}a, it is seen that the Gaussian profile's capacity bound is greater than the single-sided Lorentzian's in a region of small to medium Doppler shift, while for large Doppler shift fluctuations (where the capacity is already low), the single-sided Lorentzian slightly outperforms the Gaussian.
For small fluctuations in delay and Doppler shift (where the capacity bound is high), the Gaussian outperforms the single-sided Lorentzian significantly.

The difference between Gaussian and double-sided Lorentzian is less significant, as seen in Fig. \ref{fig:fluc_capacity_diff}b.
In the upper region of large delay fluctuations, they perform identically.
This is once again due to the independence of phase from the Doppler shift for these profiles: for delay fluctuations the capacity bound is determined by the dephasing capacity \eqref{eq:cap_deph} which for both profiles is equal and independent from the Doppler shift.
For strong Doppler fluctuations, the double-sided Lorentzian outperforms the Gaussian.
This, of course, reflects the behavior of the ambiguity function already seen in Sec. \ref{sec:amb_function}, Fig. \ref{fig:overlaps}a.

The double-sided Lorentzian also provides a superior capacity bound in comparison to the single-sided Lorentzian for a large part of the parameter range, except for relatively large Doppler shifts where the capacity is already low) as depicted in Fig.\ref{fig:fluc_capacity_diff}c.
The general features agree with Fig. \ref{fig:fluc_capacity_diff}a, since the Gaussian and double-sided Lorentzian profiles provide similar capacity bounds.

\subsubsection{Pure Doppler shifts}

\begin{figure*}
    \centering
    \includegraphics[width=.9\linewidth]{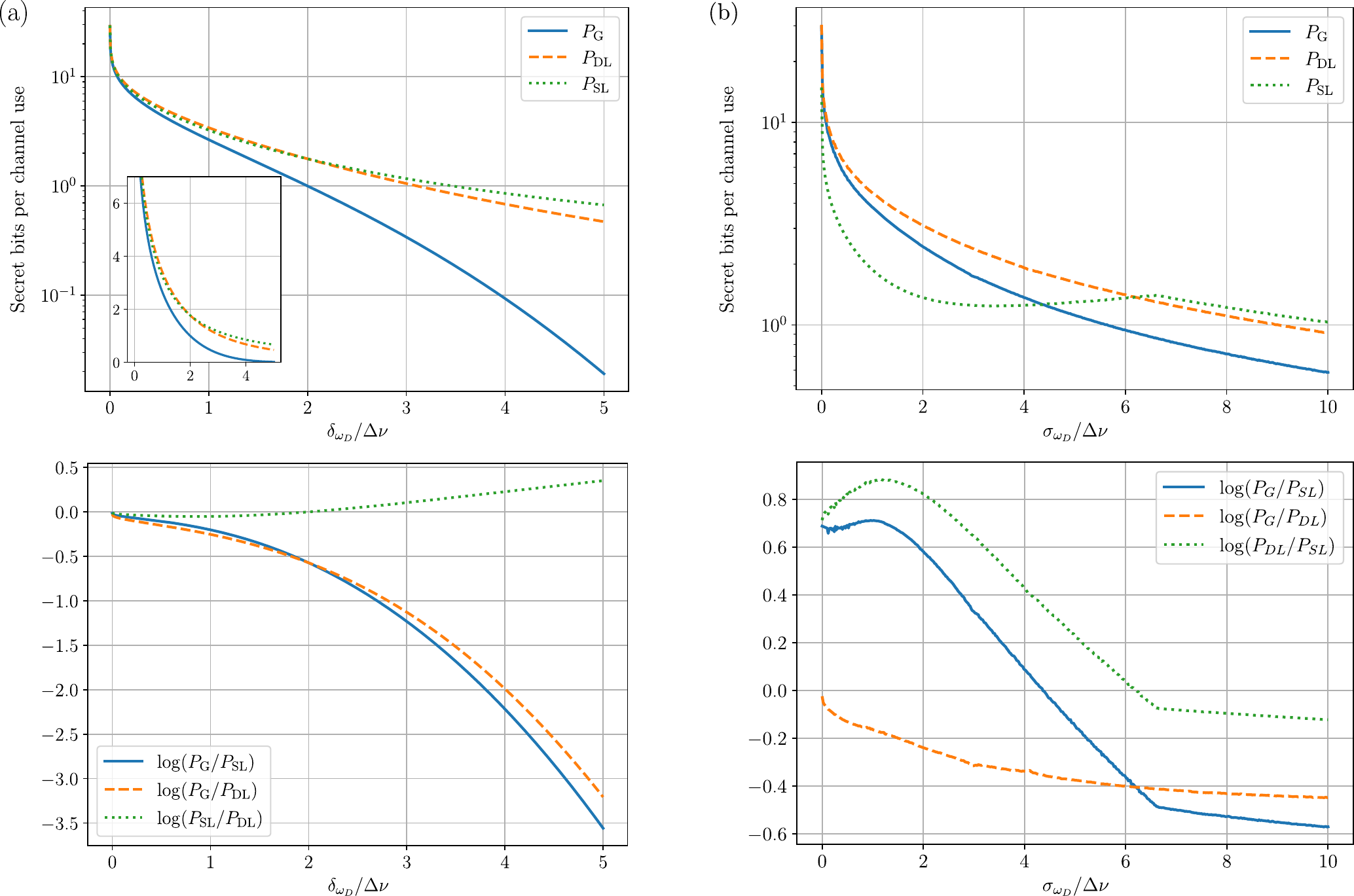}
    \caption{
    Private capacity bounds (top) and logarithms of their quotients (bottom) in dependence of (a) constant (systematic, Eq. \eqref{eq:loss_deph_capacity}) and (b) fluctuating (stochastic, Eq. \eqref{eq:fading_loss_deph_capacity}) Doppler shift and perfect delay compensation for three different spectral profiles: Gaussian (blue), double-sided Lorentzian (orange), single-sided Lorentzian (green).
    (a) The Doppler shift causes a spectral mode mismatch, leading to an effective loss that degrades the channel capacity.
    The Gaussian provides the lowest capacity, while the single-sided Lorentzian provides the highest capacity.
    (b)
    Although many features are similar to those in (a), the single-sided Lorentzian profile experiences a significant drop due to additional phase fluctuations induced by the Doppler shift, as explained in the main text.
    The other profiles do not exhibit these phase fluctuations, making the single-sided Lorentzian perform the worst up to a certain threshold.
    Due to the phase fluctuations the plot in (b) shows only an upper bound of the capacity for the single-sided Lorentzian.
    The logarithms of the quotients converge for small Doppler fluctuations, $\sigma_{\omega_D}\ll\Delta\nu$, like $\log(P_G/P_{SL}) \sim \log(P_{DL}/P_{SL})\sim\log(2)$, meaning that, for asymptotically small Doppler fluctuations, the single-sided Lorentzian yields at most half the private capacity of the other spectral profiles.
    The Gaussian and double-sided Lorentzian asymptotically yield equal capacity bounds, $\log(P_{G}/P_{DL})\sim\log(1)$.
    This can also be seen directly from the asymptotics derived in the Appendix \ref{app:cap_asymptotics}.
    }
    \label{fig:capacities}
\end{figure*}

The capacities \eqref{eq:loss_deph_capacity} for a residual Doppler shift (at zero Doppler and delay fluctuations) for the different spectral profiles are depicted in Fig. \ref{fig:capacities}a.
The capacity is a monotonically decreasing function of the residual Doppler shift $\delta_{\omega_D}$.
At larger shifts, the Lorentzian profiles have a significantly higher capacity than the Gaussian profile which is falling off exponentially.
Naturally, in the limit $\omega_D \to \infty$ all capacity bounds approach zero.
The difference between the capacity bounds of  Gaussian and Lorentzian profiles is shown explicitly in Fig. \ref{fig:capacities}a (bottom).
The differences in capacity between the profiles each show a distinct maximum which is most prominent in the difference between the Gaussian and the single-sided Lorentzian profile.
For small residual Doppler shifts, the double-sided Lorentzian provides the highest capacity bound, outperforming the other profiles.
At residual Doppler shifts of about $\omega_D \gtrsim 2 \Delta\nu$, the single-sided Lorentzian has a larger capacity than the double-sided.

We now assume that the systematic Doppler shift can be fully corrected, such that only stochastic errors in the Doppler shift remain. 
Since these Doppler fluctuations do not cause phase fluctuations for the Gaussian and double-sided spectral profiles, the quantum channel is a pure loss channel and the capacity bound is attained.
For the single-sided Lorentzian spectral profile, as discussed in Section \ref{sec:pulse_shape_optimization}, Doppler fluctuations induce phase fluctuations and the channel becomes a fading lossy dephasing channel as in \eqref{eq:fading_loss_deph_capacity}.
The corresponding capacities and the capacity bound are depicted in Fig. \ref{fig:capacities}b.
From the comparison of Fig. \ref{fig:capacities}a and \ref{fig:capacities}b we can tell that fluctuations in the Doppler shift, in contrast to a systematic error, cause a significant reduction in the capacity (bound) for the single-sided Lorentzian in comparison to the other profiles.
For larger fluctuations of about $\sigma_{\omega_D} \gtrsim 6 \Delta \nu$ the single-sided Lorentzians bound on the capacity is larger than the capacities for the Gaussian and double-sided Lorentzian spectra.
However, it should be kept in mind that, for the single-sided Lorentzian, an upper bound for the capacity, not the capacity itself, is depicted.
We conclude that the single-sided Lorentzian's performance is significantly reduced in the presence of Doppler shift fluctuations as compared to solely systematic Doppler shift.

In the applications of quantum communication that this article is relevant for, such as satellite communication links, the Doppler shift factor $z = v_{rel}/c$ is typically small, $z\ll 1$ (e.g. $z\sim 10^{-5}$ for LEO satellites \cite{ali1998doppler}).
The ratio of peak frequency to bandwidth $\frac{\omega_0}{\Delta \nu}$, however, is usually very large in quantum communication applications (e.g. $10^{8} - 10^{11}$, \cite{PhysRevLett.125.010502, frohlich2017long, zhang2019continuous, yin2017satellite2}). 
In that case, the overlap of the Gaussian spectra decays exponentially while the Lorentzians decay only hyperbolically.
Analytic expressions for the overlap asymptotics are given in the Appendix \ref{app:asymptotics}.

\subsubsection{Pure delays}
If we consider a systematic error in arrival time (delay), we assume the phase error to be measurable and corrected in post-processing, as explained in \ref{sec:quantumchannelcapacity}. Although the phase can be corrected, the delay still causes an effective loss due to the non-unity overlap of the wave packets. This corresponds to the modulus of the ambiguity function, which was already discussed in Section \ref{sec:amb_function} and Fig. \ref{fig:overlaps}b.
The quantum capacity is then simply given by the PLOB bound \eqref{eq:cap_plob}. 

For stochastic errors, that is,
fluctuating delays, in the absence of Doppler shift, all spectral profiles show the same behavior in terms of capacity. The capacity is upper bounded by the dephasing channel capacity \eqref{eq:cap_deph} which is independent of the spectral profile in question since the phase of the ambiguity function reduces to $-\omega_0 \tau$ in this case (see also Section \ref{sec:amb_function}).

\section{Conclusion} \label{sec:conclusion}
We have investigated the resilience of optical quantum communication systems to spatio-temporal distortions.
In the context of continuous-variable quantum key distribution protocols, our findings reveal a direct connection between the behavior of generalized correlation functions and channel capacities, emphasizing the fundamental relationship between cryptographic security and spectral characteristics.

Further analysis centered on Gaussian and Lorentzian spectral profiles, representing distinct paradigms in spectral engineering.
Our results demonstrate the superior ability of Gaussian profiles to mitigate the detrimental effect of temporal delays compared to Lorentzian profiles, whereas Lorentzian profiles effectively reduce the vulnerability to spectral deformations originating from Doppler shifts.
We find, however, that in the presence of fluctuating Doppler shift, the single-sided Lorentzian profile can lead to significant dephasing, making it less preferable in such scenarios.
This suggests that optimizing spectral characteristics could significantly improve system resilience.

The practical implications of our findings extend beyond theoretical insights, offering concrete guidance for the design of optical quantum communication systems.
We believe that the addition of spectral optimization to compensation techniques will yield a more universally resilient design of optical quantum network infrastructure.

To better understand the implications of our results for practical implementations of space-based quantum communication, the present analysis has to be extended to include realistic satellite and ground station constellations and a model of the timing of emission and reception of signals (the emitter-observer problem), which will be addressed in a follow-up article.

Beyond leading order Doppler shifts, the significantly smaller relativistic contributions of transverse Doppler shift and gravitational redshift are also covered by our analysis as they affect the signal by frequency shifts only.
However, the inclusion of higher order relativistic contributions necessitates a more detailed modeling.
For example, to include spatial mode properties and how they are affected by gravity in a relativistic situation, our analysis has to be extended to 3-dimensional optical pulses.
Such a framework would be able to cover effects like wavefront deformations due to spacetime curvature and the Wigner translation of polarized light due to the relative motion of emitter and observer.
Eventually, this should provide a general framework to investigate the resilience of quantum communication networks to relativistic effects, which can also be employed to quantify the utility of such networks for tests of general relativity and quantum optics in curved spacetime \cite{Zych_2012,Rideout_2012, deep_space_quantum_link, BELENCHIA20221_Quantum_Physics_in_Space,  Mieling_2022_Measuring_spacetime_curvature, bruschi2014spacetime, kohlrus2017quantum, bruschi2021spacetime, barzelPB, Barzel2024entanglement, borregaardTestingQuantumTheory2024}.

\section*{Acknowledgement}
We gratefully acknowledge Mustafa Gündoğan, David Edward Bruschi, and Andreas Wolfgang Schell for fruitful discussions on the topic. We also want to thank Dennis Phillip and Volker Perlick for their useful comments and discussions.
E.S. and D.R. acknowledge funding by the Federal Ministry of Education and Research of Germany in the project “Open6GHub” (grant number: 16KISK016).
R.B. was funded by the Deutsche Forschungsgemeinschaft (DFG, German Research Foundation) under Germany’s Excellence Strategy – EXC-2123 QuantumFrontiers – 390837967.
We further gratefully acknowledge funding through the CRC TerraQ from the Deutsche Forschungsgemeinschaft (DFG, German Research Foundation) – Project-ID 434617780 – SFB 1464 and the Research Training Group 1620 “Models of Gravity”.

\appendix

\section{Derivation of the generalized correlation function
\label{app:amb_fun}}

For the spectral amplitudes of interest (Gaussian and single- and double-sided Lorentzians) we can analytically calculate the ambiguity function as given in Eq. \eqref{eq:amb_function}.

\subsection{Gaussian}
To calculate the ambiguity function for the Gaussian spectrum we integrate in the spectral domain and make use of the fact that the product of two Gaussians is a Gaussian and that the Fourier transform of a Gaussian is also Gaussian.
\begin{equation}
\begin{aligned}
    &Q_G(z,\tau) = \sqrt{1+z} \int F^*_G(\omega) F_G(\omega(1+z)) e^{ - i \omega \tau} d\omega
    \\
    &=\sqrt{\frac{1+z}{2\pi \sigma^2}} \int
    e^{-\frac{1}{4 \sigma^2}\left(
    2(\omega-\omega_0)^2 + 2(\omega-\omega_0)\omega z + \omega^2z^2
    \right)} e^{- i\omega \tau} d\omega
    \\
    &=\sqrt{\frac{1+z}{2\pi \sigma^2}} \int
    e^{-\frac{1}{4 \sigma^2}\left(
    \omega^2(1+(1+z)^2) - \omega\omega_02(2+z) +2 \omega_0^2
    \right)} e^{- i\omega \tau} d\omega
    \\&=
    \sqrt{\frac{2(z+1)}{z (z+2)+2}} e^{ -\frac{\omega_0^2 z^2 + 4 i \omega_0 \sigma^2 \tau (z+2)+4 \sigma^4 \tau^2}{4 \sigma^2 (z (z+2)+2)}},
\end{aligned}
\end{equation}
where the standard deviation $\sigma$ is related to the bandwidth (here HWHM) by $\sigma = \frac{\Delta\nu}{\sqrt{\ln {4}}}$. For small Doppler shifts $z\ll 1$ while $\omega_D:= z \omega_0$ remains significant, we can retrieve the Woodwards ambiguity function in the narrowband limit
\begin{equation}
    \chi_G(\omega_D, \tau) = e^{-\frac{\omega_D^2}{8 \sigma ^2}-\frac{1}{2} \left(\sigma ^2 \tau^2\right)} 
    e^{- i \tau \left(\omega_0 - \frac{\omega_D}{2}\right)
    }.
\end{equation}

\subsection{Double-sided Lorentzian}
The calculation for the double-sided Lorentzian can be performed either in the spectral domain by making use of the residual theorem or by direct integration of the exponentials in the temporal domain. We give the latter derivation.
The spectral amplitude function in the temporal domain is the double-sided exponential function
\[
A_{DL}(t) = \sqrt{s} e^{-s |t| + i\omega_0 t},
\]
where $s$ is the width parameter of the Lorentzian and related to the bandwidth by $\Delta \nu = s \sqrt{\sqrt{2}-1}$.
The ambiguity function is then given as
\begin{align}
    Q_{DL}(z,\tau) = s\sqrt{1+z} \int_{-\infty}^\infty e^{-s(|t| + |\tau+t(1+z)|} e^{-i\omega_0 (\tau+tz)} dt.
\end{align}
Because the absolute value of $\tau$ enters into the expression we consider each sign separately.
The overall ambiguity function is then the composition
\begin{equation}
    Q_{DL}(z,\tau) = s\sqrt{1+z}
    \begin{cases} q^>_{DL}(z,\tau),&\tau\geq 0,
    \\ q^<_{DL}(z,\tau),& \tau< 0.
    \end{cases} 
\end{equation}

For $\tau>0$ we integrate three intervals.
\begin{enumerate}
\item $I = \{ t>0 \}$
\begin{equation}
\begin{aligned}
    q^>_I(z,\tau) &= \int_0^\infty e^{-s(\tau+t(2+z)} e^{-i\omega_0 (\tau+tz)} dt
    \\
    &=\frac{e^{-s\tau-i\omega_0\tau}}{s(2+z) + i\omega_0 z}
\end{aligned}
\end{equation}
\item $II = \{t<0 \land t> -\tau/(1+z) \}$
\begin{equation}
    \begin{aligned}
        q^>_{II}(z,\tau) &= \int_{-\frac{\tau}{z+1}}^{0} e^{-s(\tau+tz)} e^{-i\omega_0 (\tau+tz)} dt
        \\& = \frac{e^{-s \tau - i\omega_0 \tau}}{s z + i\omega_0 z}
        \left( e^{s \tau \frac{z}{z+1} + i\omega_0 \tau \frac{z}{z+1}}
        - 1\right)
    \end{aligned}
\end{equation}
\item $III = \{ t< \tau/(1+z) \}$
\begin{equation}
    \begin{aligned}
        q^>_{III}(z,\tau) &= \int_{-\infty}^{\frac{-\tau}{z+1}}
        e^{s (\tau + t(2+z)} e^{-i\omega_0 (\tau + tz)} dt
        \\&=
        \frac{e^{s \tau -i\omega_0 \tau}}{s (2+z) - i\omega_0 z}
        e^{-s \tau \frac{2+z}{1+z} + i \omega_0 \tau \frac{z}{z+1}}
        \\&=
        \frac{e^{-s \tau -i\omega_0 \tau}}{s (2+z) - i\omega_0 z}
        e^{s \tau \frac{z}{z+1} + i\omega_0 \tau \frac{z}{z+1}}
    \end{aligned}
\end{equation}
\end{enumerate}

The sum of all three terms yields for $\tau>0$
\begin{equation}
\begin{split}
    q^>_{DL}(z,\tau)
    = &e^{\frac{-\tau(s+i\omega_0)}{z+1}} \left(
    \frac{1}{s z + i\omega_0 z} + \frac{1}{s(z+2) - i \omega_0 z}
    \right)
    \\ -&e^{-\tau(s + i\omega_0)}\left(
    \frac{1}{s z + i \omega_0 z} - \frac{1}{s(2+z) + i \omega_0 z} 
    \right)
\end{split}
\end{equation}

For $\tau<0$ an analogous calculation yields
\begin{equation}
\begin{split}
    q^<_{DL}(z,\tau)
    = &e^{\frac{\tau(s-i\omega_0)}{z+1}} \left(
    \frac{1}{s z - i\omega_0 z} + \frac{1}{s(z+2) + i \omega_0 z}
    \right)
    \\ -&e^{\tau(s - i\omega_0)}\left(
    \frac{1}{s z - i \omega_0 z} - \frac{1}{s(2+z) - i \omega_0 z} 
    \right).
\end{split}
\end{equation}

For small Doppler shifts $z\ll 1$ we can make the following approximation
\begin{equation}
\begin{aligned}
    \chi_{DL}(z,\tau)& = e^{-s |\tau| 
    -i\tau \left( \omega_0 -\frac{1}{2}\omega_D \right) }
    \frac{
    \cos\left(\frac{\omega_D |\tau|}{2}\right) + \frac{2s}{\omega_D} \sin\left(\frac{\omega_D |\tau|}{2}\right)
    }{1+\left(\frac{\omega_D}{2 s}\right)^2}.
\end{aligned}
\end{equation}

\subsection{Single-sided Lorentzian}
The ambiguity function for the single-sided exponential function can be found by straightforward integration
\begin{equation}
\begin{aligned}
    \begin{aligned}
	Q_{SL}(z,\tau) =& 2\Delta\nu \sqrt{1+z} \int e^{-\Delta \nu (t(2+z) + \tau)} e^{-i\omega_0 (\tau+tz)}
    \\ &\qquad \times \Theta(t) \Theta\left(t(1+z)+\tau\right) dt
    \end{aligned}
    \\
    \begin{aligned}
    =& 2\Delta\nu \sqrt{1+z} \bigg(
    \int_0^\infty e^{-\Delta \nu (t(2+z) + \tau)} e^{-i\omega_0 (\tau+tz)}
    \Theta(\tau) dt
    \\
    +&\int_{-\tau/(1+z)}^\infty e^{-\Delta \nu (t(2+z) + \tau)} e^{-i\omega_0 (\tau+tz)}
    \Theta(-\tau) dt
    \bigg)
    \end{aligned}
    \\
    \begin{aligned}
    = &2 \Delta\nu \sqrt{1+z}\frac{
    {e^{-\tau(\Delta\nu + i\omega_0)}}
    \Theta(\tau)
    +e^{\frac{\tau}{1+z} (\Delta\nu - i \omega_0)} \Theta(-\tau)
    }{\Delta \nu (z+2) + i \omega_0 z}
    \end{aligned}
\end{aligned}
\end{equation}

We again approximate the above expression for small Doppler shifts $z\ll 1$
\begin{align}
	\chi_{SL}(\omega_D, \tau) =
    \frac{e^{-|\tau| \Delta \nu}}{1 + i\frac{\omega_D}{2\Delta\nu}}
    \begin{cases}
    {e^{-i\tau (\omega_0-\omega_D)}}
    ,& \tau <0
    \\
    {e^{-i\tau \omega_0}}
    ,& \tau \geq 0.
    \end{cases}
\end{align}

\section{Asymptotics}
\label{app:asymptotics}

\subsection{Ambiguity functions}
\label{app:amb_asymptotics}

Here we give the expressions for the asymptotics of the ambiguity functions \eqref{eq:wamb_gauss},\eqref{eq:wamb_dl},\eqref{eq:wamb_sl}
for $\omega_D/\Delta\nu \ll 1$
\begin{align}
&\chi_G(\omega_D,\tau) \sim \left(1-\frac{1}{2}\left(\frac{\omega_D}{2 \sigma}\right)^2 \right) e^{-i(\omega_0-\frac{\omega_D}{2}) \tau -\frac{\tau^2\sigma^2}{2}},\\
&\begin{aligned}
\chi_{DL}(\omega_D,\tau) &\sim \left(1-\left(\frac{\omega_D}{2 s}\right)^2 \right) e^{-i(\omega_0 - \frac{\omega_D}{2})\tau - s| \tau|}
\\
\quad\times &\left(
\cos\left(\frac{\omega_D|\tau|}{2}\right)+ \frac{2s}{\omega_D}\sin\left(\frac{\omega_D|\tau|}{2}\right)
\right),
\end{aligned}
\\
&\begin{aligned}
		\chi_{SL}(\omega_D,\tau) &\sim \left( 1- \frac{1}{2}
\left(\frac{\omega_D}{2 \Delta\nu}\right)^2 \right) \\
\quad\times& e^{-i  \frac{\omega_D}{2\Delta\nu} -\Delta\nu |\tau|}
\begin{cases}
    e^{- i (\omega_0 - \omega_D) \tau},& \tau<0
    \\ e^{-i\omega_0 \tau},& \tau \geq 0
\end{cases}
\end{aligned}
,
\end{align} 
 and for $\omega_D/\Delta\nu \gg 1$
\begin{align}
&\chi_G(\omega_D,\tau) \sim e^{-\frac{1}{2}\left(\frac{\omega_D}{2 \sigma}\right)^2-\frac{\tau^2\sigma^2}{2}}  e^{-i(\omega_0 - \frac{\omega_D}{2}) \tau},\\
&\chi_{DL}(\omega_D,\tau) \sim \left(\frac{\omega_D}{2 s}\right)^{-2}  e^{-s|\tau| - i (\omega_0-\frac{\omega_D}{2})\tau}
\cos\left(\frac{\omega_D|\tau|}{2}\right)
,
\\
&\begin{aligned}
\chi_{SL}(\omega_D,\tau) &\sim \left(\frac{\omega_D}{2 \Delta\nu}\right)^{-1} e^{- \Delta\nu|\tau|}
\\
\quad\times&e^{- i\pi/2}
\begin{cases}
    e^{- i (\omega_0 - \omega_D) \tau},& \tau<0
    \\ e^{-i\omega_0 \tau},& \tau \geq 0.
\end{cases}
\end{aligned}
\end{align}

\subsection{Capacities \label{app:cap_asymptotics}}

Here we give the asymptotics of the capacities as obtained from the ambiguity functions' asymptotics derived above.
\subsubsection{Systematic Doppler shift}
In the presence of systematic Doppler shift and zero delay, the capacity is determined by the PLOB bound \eqref{eq:cap_plob}.
Using the asymptotics for the ambiguity function, we obtain for $\delta_{\omega_D}/\Delta\nu\ll 1$
\begin{align}
P_G(\delta_{\omega_D}) &\sim 
-\log_2 \left(\frac{1}{2}\left(\frac{\delta_{\omega_D}}{2\sigma}\right)^2\right)
\\
P_{DL}(\delta_{\omega_D}) &\sim
-\log_2 \left(\frac{\delta_{\omega_D}}{2 s}\right)^2
\\
P_{SL}(\delta_{\delta_{\omega_D}}) &\sim 
-\log_2 \left(\frac{1}{2}\left(\frac{\delta_{\omega_D}}{2 \Delta \nu}\right)^2\right).
\end{align}
\subsubsection{Stochastic Doppler shift}
For an asymptotically small stochastic Doppler shift, $\sigma_{\omega_D}/\Delta\nu \ll 1$, and zero delay the capacities for the Gaussian and single-sided Lorentzian profile are determined by the averaged PLOB bound \eqref{eq:fading_loss_plob}, yielding
\begin{align}
    P_G(\sigma_{\omega_D}) &\sim
    -\log_2 \left(\frac{1}{4}\left(\frac{\sigma_{\omega_D}}{2 \sigma}\right)^2\right)
    +\frac{\gamma}{\log 2}
    \\
    P_{DL}(\sigma_{\omega_D}) &\sim
    -\log_2 \left(\frac{1}{2}\left(\frac{\sigma_{\omega_D}}{2 s}\right)^2\right)
    +\frac{\gamma}{\log 2},
\end{align} where $\gamma$ is the Euler–Mascheroni constant,
meaning both agree asymptotically, $P_G \sim P_{DL}$.
Since the single-sided Lorentzian suffers dephasing from Doppler fluctuations, the corresponding capacity is upper bounded by the minimum of PLOB bound and dephasing capacity, see \eqref{eq:fading_loss_deph_capacity}. The asymptotic PLOB bound is
\begin{align}
    P_{SL}(\mathcal{L}_\eta)=
    -\log_2 \left(\frac{1}{4}\left(\frac{\sigma_{\omega_D}}{2 \Delta\nu}\right)^2\right)
    +\frac{\gamma}{\log 2},
\end{align}
while the asymptotic dephasing capacity is only \cite{lamiExactSolutionQuantum2023}
\begin{align}
    P_{SL}(\mathcal{N}_p)=
    -\frac{1}{2} \log_2 \left(\frac{e }{2 \pi}
    \left(\frac{\sigma_{\omega_D}}{2\Delta \nu}\right)^2\right).
\end{align}
Therefore, asymptotically, the dephasing bounds the capacity for the single-sided Lorentzian profile to half the capacity of the remaining channels which remain pure loss, i.e. $P_{SL} \sim \frac{1}{2} P_G \sim \frac{1}{2} P_{DL}$.

\bibliographystyle{apsrev4-2}
\bibliography{bib}

\end{document}